\documentclass[12pt]{iopart}

\expandafter\let\csname equation*\endcsname\relax
\expandafter\let\csname endequation*\endcsname\relax

\usepackage{amsmath}
\usepackage{esint}
\usepackage{amssymb}
\usepackage{mathtools}
\usepackage{bm}
\usepackage{setspace}

\usepackage{csquotes}
\usepackage[backend=biber, style=numeric, style=science, sorting=none, doi=true, url=false, citestyle=numeric-comp, eprint=true]{biblatex}
\usepackage[
     unicode=true,
     bookmarks=true,
     bookmarksnumbered=true,
     bookmarksopen=true,
     bookmarksopenlevel=3,
     breaklinks=true,
     pdfborder={0 0 1},
     backref=false]
     {hyperref}
\hypersetup{
    colorlinks=true,
    linkcolor=blue,
    anchorcolor=.,
    citecolor=blue,
    filecolor=.,
    menucolor=.,
    runcolor=.,
    urlcolor=blue}

\begin{document}

\title[]{Simulations of edge and SOL turbulence in diverted negative and positive triangularity plasmas
}

\author{P. Ulbl$^{1}$, A. Stegmeir$^1$, D. Told$^1$, G. Merlo$^1$, K. Zhang$^1$ and F. Jenko$^{1,2}$}

\address{$^1$ Max-Planck-Institute for Plasma Physics, Boltzmannstrasse 2, 85748 Garching, Germany}
\address{$^2$ University of Texas at Austin, Austin, Texas 78712, USA}
\ead{philipp.ulbl@ipp.mpg.de}
\vspace{10pt}
\begin{indented}
\item[]\today
\end{indented}

\begin{abstract}
Optimizing the performance of magnetic confinement fusion devices is critical to achieving an attractive fusion reactor design. Negative triangularity (NT) scenarios have been shown to achieve excellent levels of energy confinement, while avoiding edge localized modes (ELMs). Modeling turbulent transport in the edge and SOL is key in understanding the impact of NT on turbulence and extrapolating the results to future devices and regimes. Previous gyrokinetic turbulence studies have reported beneficial effects of NT across a broad range of parameters. However, most simulations have focused on the inner plasma region, neglecting the impact of NT on the outermost edge. In this work, we investigate the effect of NT in edge and scrape-off layer (SOL) simulations, including the magnetic X-point and separatrix. For the first time, we employ a multi-fidelity approach, combining global, non-linear gyrokinetic simulations with drift-reduced fluid simulations, to gain a deeper understanding of the underlying physics at play. First-principles simulations using the \texttt{GENE-X} code demonstrate that in comparable NT and PT geometries, similar profiles are achieved, while the turbulent heat flux is reduced by more than 50\% in NT. Comparisons with results from the drift-reduced fluid turbulence code \texttt{GRILLIX} suggest that the turbulence is driven by trapped electron modes (TEMs). The parallel heat flux width on the divertor targets is reduced in NT, primarily due to a lower spreading factor $S$.


\end{abstract}

\section{Introduction}

A key challenge in magnetic confinement fusion research is achieving optimal energy confinement while maintaining heat and particle exhaust within acceptable limits for plasma-facing components. In this context, plasma shaping, particularly negative triangularity (NT), has been a major focus of research in recent years \cite{Marinoni2021a}. Compared to its positive triangularity (PT) counterpart, NT has demonstrated favorable energy confinement \cite{Camenen2007} while naturally suppressing edge localized modes (ELMs) \cite{Nelson2023}. NT plasmas have been studied in various experimental devices, including TCV~\cite{Pochelon1999, Coda2022}, \mbox{DIII-D}~\cite{Austin2019, Marinoni2021b, Thome2024}, AUG~\cite{Happel2023}, and JET~\cite{Labit2024}. Recent studies have predicted high-confinement NT regimes in DIII-D \cite{Marinoni2024} and DTT~\cite{Mariani2024}, and a conceptual design for a fusion pilot plant based on NT has been developed \cite{MANTA2024}.

To extrapolate the behavior of NT plasmas from existing experiments to next-step devices, a deep theoretical understanding of the underlying physics is crucial. Global gyrokinetic turbulence simulations provide the most accurate approach for studying turbulent transport, but their computational cost is significant. Reduced models, such as local gyrokinetic simulations or global drift-reduced fluid computations, offer more efficient alternatives. However, to ensure their reliability, it is essential to verify their results through cross-comparisons with each other and with higher-fidelity models. A multi-fidelity approach enhances our understanding of the underlying physics by systematically comparing simulations based on different modeling assumptions. In this work, we take a first step in this direction by comparing global nonlinear gyrokinetic simulations with global nonlinear drift-reduced fluid simulations.


Up to now, gyrokinetic simulations of turbulent transport in NT plasmas have primarily focused on the core plasma region. Local flux-tube simulations have shown that NT has a beneficial effect on trapped electron mode (TEM) turbulence \cite{Marinoni2009,Marinoni2021b, Merlo2023a} and ion temperature gradient (ITG) turbulence \cite{Merlo2023a, Merlo2023b, Balestri2024}. The impact of aspect ratio has been investigated \cite{Marinoni2024, Balestri2024}. Global simulations within the confined plasma region have been carried out using delta-$f$ gyrokinetic models for TCV \cite{Merlo2021, DiGiannatale2024} and gyrokinetic/gyromoment models in DIII-D \cite{Li2024, Hoffmann2025}. Global full-$f$ gyrokinetic simulations, including a limiter, have been performed for DIII-D \cite{Bernard2024}. The first non-linear full-$f$ simulations in NT-diverted geometry, including the magnetic X-point, were presented in Refs. \cite{UlblIAEA,Poli2024}. This study builds upon and extends that work by employing a more accurate physics model. In particular, the improved collision operator employed here (see Section \ref{sec:genex_model} for details) was informed by a code validation study on TEM-dominated turbulence in TCV \cite{Ulbl2023}. Additionally, recent non-linear global gyrokinetic simulations have been carried out for TCV and DIII-D \cite{Becoulet2024}. The present work aims at significantly extending these studies.

The remainder of this work is structured as follows. In Section 2, we introduce the models and methods used in the present multi-fidelity study, including non-linear global full-$f$ gyrokinetic and drift-fluid simulations, as well as local delta-$f$ flux-tube simulations. In Section 3, we describe the simulation setup necessary to reproduce the presented results. Section 4 details the first-principles simulation results on profiles, heat fluxes, energy confinement, turbulence, and divertor heat fluxes. In Section 5, we summarize the key findings and provide an outlook on future research directions.

\section{Models and methods}

In this section, we provide a brief overview of the physics models employed in this study and introduce the numerical codes used to solve the governing equations. Specifically, we will utilize the \texttt{GENE-X} \cite{Michels2021} and \texttt{GRILLIX} \cite{Stegmeir2018} codes to perform global turbulence simulations in X-point geometry. The differing fidelities of these two physics models will enable us to partially isolate physics effects in the simulations conducted in this work. Additionally, we will employ the fluxtube version of the \texttt{GENE} code \cite{Jenko2000} to identify the linearly dominant micro-instabilities.

\subsection{Global gyrokinetic simulations in X-point geometry}\label{sec:genex_model}

The primary tool employed in this work is the \texttt{GENE-X} code \cite{Michels2021}, which utilizes the same physics model as described in Ref. \cite{Ulbl2023}. For convenience, we provide a brief summary of the physics model below.

The \texttt{GENE-X} code solves the full-$f$ gyrokinetic Vlasov-Maxwell system, including electromagnetic effects and collisions. The five-dimensional gyrocenter distribution function of plasma species $\alpha$ is represented as $f_\alpha(\mathbf{x}, v_{||}, \mu)$, where the coordinates denote the gyrocenter position, parallel velocity, and magnetic moment, respectively. The evolution of the distribution function in time t is governed by the gyrokinetic equation
\begin{align}
    \frac{\partial f_\alpha}{\partial t}
    &+ v_{||} \, \frac{\mathbf{B}^*}{B_{||}^*} \cdot \nabla f_\alpha + \frac{c}{q_\alpha B_{||}^*} \, \mathbf{b}\times (\mu\nabla B + q_\alpha \nabla \phi_1) \cdot \nabla f_\alpha \nonumber\\
    &- \frac{\mathbf{B}^*}{m_\alpha B_{||}^*} \cdot (\mu \nabla B + q_\alpha \nabla \phi_1) \, \frac{\partial f_\alpha}{\partial v_{||}} - \frac{q_\alpha}{m_\alpha c} \frac{\partial A_{1,||}}{\partial t} \,\frac{\partial f_\alpha}{\partial v_{||}} = C_\alpha f_\alpha,\label{eq:gyrokin}
\end{align}
with the guiding-center magnetic field $\mathbf{B}^* = \mathbf{B} + (m_\alpha c/q_\alpha) v_{||} \nabla \times \mathbf{b} + \nabla A_{1,\|} \times \mathbf{b}$ and $B_{||}^* = \mathbf{b}\cdot\mathbf{B}^*$. Here $q_\alpha$ and $m_\alpha$ represent the species charge and mass, respectively, $c$ is the speed of light and $\mathbf{B}$ is the background magnetic field with unit vector $\mathbf{b}$ and magnitude $B$.

The electrostatic potential $\phi_1$ and parallel electromagnetic potential $A_{1,||}$ are allowed to fluctuate in time and are determined self-consistently through the field equations
\begin{align}
    -\nabla \cdot \left(\sum_\alpha \frac{m_\alpha c^2 n_{0,\alpha}}{B^2} \nabla_\perp \phi_1\right) &= \sum_\alpha q_\alpha \int f_\alpha \,\mathrm{d}V,\label{eq:qn}\\
    - \Delta_\perp A_{1,||} &=4\pi \sum_\alpha \frac{q_\alpha}{c} \int v_{||} f_\alpha \,\mathrm{d}V.\label{eq:ampere}
\end{align}
These equations are known as the quasi-neutrality (QN) equation and Ampere's law, respectively. The right-hand sides of both equations represent the gyrocenter charge and current densities, which are obtained through velocity space integrals (moments) in $\mathrm{d}V=2\pi B_{||}^*/m_\alpha\, \mathrm{d}v_{||}\,\mathrm{d}\mu$. The left hand side differential operators are the perpendicular gradient $\nabla_\perp = (\bm{\mathrm{I}} - \mathbf{b}\mathbf{b}) \cdot \nabla$ and perpendicular Laplacian $\Delta_\perp = \nabla \cdot \nabla_\perp$. The index 0 denotes the initial $t=0$ gyrocenter density used in the QN equation.

The right hand side of the gyrokinetic equation (\ref{eq:gyrokin}) models the effects of Coulomb collisions using a bi-linear collision operator $C_\alpha f_\alpha=\sum_\beta C_{\alpha\beta}(f_\alpha, f_\beta)$. Specifically, the Lenard-Bernstein/Dougherty (LBD) collision operator is employed in this work,
\begin{align}
    C_{\alpha\beta}^\mathrm{LBD} f_\alpha = \frac{\nu_{\alpha\beta}}{B_{||}^*} \Bigg\{
                        &\frac{\partial}{\partial v_{||}} \left[
                            \left(v_{||} - u_{\alpha\beta}\right) B_{||}^* f_\alpha + \frac{T_{\alpha\beta}}{m_\alpha}
                                \frac{\partial B_{||}^* f_\alpha}{\partial v_{||}}\right]\nonumber\\
                        +&\frac{\partial}{\partial \mu} \left[
                            2 \mu B_{||}^* f_\alpha + \frac{2 T_{\alpha\beta}}{B} \mu
                                \frac{\partial B_{||}^* f_\alpha}{\partial \mu}\right] \Bigg\}.\label{eq:lbd}
\end{align}
This collision operator is of Fokker-Planck type, with simplified friction and diffusion coefficients proportional to so-called \enquote{mixing} flows $u_{\alpha\beta}$ and temperatures $T_{\alpha\beta}$. These are defined as
\begin{align}
u_{\alpha\beta} &= \frac{\nu_{\alpha\beta} m_{\alpha} n_{\alpha} u_{\alpha} + \nu_{\beta\alpha} m_{\beta} n_{\beta} u_{\beta}}{\nu_{\alpha\beta} m_{\alpha} n_{\alpha} + \nu_{\beta\alpha} m_{\beta} n_{\beta}},\label{eq:u_mix}\\
T_{\alpha\beta}
&= \frac{T_{\alpha} \nu_{\alpha\beta} n_{\alpha} + T_{\beta} \nu_{\beta\alpha} n_{\beta}}{\nu_{\alpha\beta} n_{\alpha}+ \nu_{\alpha\beta} n_{\alpha}} - \frac{1}{3} \frac{\nu_{\alpha\beta} n_{\alpha} m_{\alpha} \left(u_{\alpha\beta}^2-u_{\alpha}^2\right) + \nu_{\beta\alpha} n_{\beta} m_{\beta} \left(u_{\alpha\beta}^2-u_{\beta}^2\right)}{\nu_{\alpha\beta} n_{\alpha} + \nu_{\beta\alpha} n_{\beta}},\label{eq:T_mix}
\end{align}
which are linear combinations of densities, flows and moving-frame temperatures
\begin{align}
n_{\alpha} &= \int f_{\alpha} \mathrm{d}V, \\
u_{\alpha} &= \frac{1}{n_{\alpha}} \int  f_{\alpha} v_{||} \mathrm{d}V, \\
T_{\alpha} &= \frac{2}{3 n_\alpha} \int f_{\alpha} \left(\frac{m_{\alpha} v_{||}^2}{2} + \mu B \right) \mathrm{d}V  - \frac{1}{3} m_{\alpha} u_{\alpha}^2.\label{eq:T_mom}
\end{align}
using collision frequencies
\begin{align}
    \nu_{\alpha\beta} &= \frac{8 \sqrt{2\pi} n_{\beta} \sqrt{m_{\alpha} m_{\beta}} (Z_{\alpha} Z_{\beta} e^2)^2 \ln\Lambda_{\alpha\beta}}{3 (m_{\alpha} T_{\beta} + m_{\beta} T_{\alpha})^{3/2}}. \label{eq:nu}
\end{align}
Here $e$ denotes the elementary charge, $Z_\alpha$ the species charge state and $\ln \Lambda_{\alpha\beta}$ the Coulomb logarithm. We note that Refs. \cite{Ulbl2021,Ulbl2023} contain a small typo in the denominator of the second term in equation (\ref{eq:T_mix}), where species indices where inadvertently switched.

Equations (\ref{eq:gyrokin}-\ref{eq:nu}) form the system that is used in the \texttt{GENE-X} code, which employs the standard gyrokinetic ordering \cite{Brizard2007}. The induction term $\partial A_{1,||}/\partial t$ is solved by introducing a generalized Ohm's law \cite{Mandell2020}. Further details on the collisionless, electromagnetic model can be found in Refs. \cite{Michels2021, Michels2022}. The treatment of collisions is described in Refs. \cite{Ulbl2021, Ulbl2023}.

While the model is considered high-fidelity, certain approximations were made due to the complex nature of the problem. The most relevant ones for this work are as follows. A long-wavelength approximation is employed, where potentials evaluated at particle position $\mathbf{x}'$ are approximated as $\phi_1(\mathbf{x}')\approx(1+\boldsymbol{\rho}_L\cdot\nabla)\phi_1(\mathbf{x})$, where $\boldsymbol{\rho}_L$ denotes the gyrocenter displacement. This simplification limits the model to contain only leading-order finite ion Larmor radius (FLR) effects, targeting the applicability to long-wavelength $\rho_L k_\perp < 1$ turbulence in the edge and SOL.  Additionally, the collision operator (\ref{eq:lbd}) is simplified by neglecting the velocity space dependency of collision frequencies and approximating the pitch-angle scattering part. Furthermore, no neutral gas model is currently implemented, and effects of the plasma sheath in front of the divertor plates are not included.

\texttt{GENE-X} employs the flux-coordinate independent (FCI) approach \cite{Hariri2013, Stegmeir2016} to enable simulations that include the magnetic X-point and the separatrix. The discretization of the governing equations is performed using finite differences for the collisionless part \cite{Michels2021} and finite volume for the collision operator \cite{Ulbl2021}. The field equations are solved using a GMRES algorithm \cite{Saad1986} with a geometric multigrid preconditioner. The system is evolved in time using Strang splitting \cite{Strang1968} in the collisionless and collisional parts, with RK4 time integration used in each part. In this work, we utilize the improved perpendicular velocity space discretization of the collision operator given in Ref. \cite{UlblPhd}, using the perpendicular velocity $v_\perp = \sqrt{2 B \mu / m_\alpha}$ as a coordinate. Velocity space moments are calculated using the composite midpoint formula in both coordinates \cite{NumericalRecipes}. Dirichlet boundary conditions are applied on both the inner core and the wall and divertor target boundaries. The fluctuating potentials are pinned to zero, while the distribution functions of each species are fixed to Maxwellians with density and temperature given by the initial profiles. Additionally, numerical diffusion is added in a buffer region close to the real space boundary regions. Fourth-order hyperdiffusion is applied globally to suppress spurious oscillations created by the discretization \cite{Pueschel2010}. Further details on the FCI approach can be found in Ref. \cite{Michels2021}, on discretized collisions in Ref. \cite{Ulbl2021}, on numerical diffusion and boundary conditions in Ref. \cite{MichelsPhd}, and on the splitting scheme in Ref. \cite{Ulbl2023}.

\subsection{Global drift-reduced Braginskii fluid simulations in X-point geometry}

We employ the \texttt{GRILLIX} code to simulate fluid turbulence, utilizing a drift-reduced full-$f$ electromagnetic Braginskii \cite{Braginskii1965} model with trans-collisional extensions, as summarized in Appendix A of Ref.~\cite{Zholobenko2024}. The time evolution of fluid-like quantities, including density, parallel flow and current, species temperatures, and electrostatic and electromagnetic potentials, is solved directly. The fluid hierarchy is truncated at the parallel heat fluxes, where a heat-flux-limiting Braginskii expression is applied, as detailed in Ref.~\cite{Zholobenko2021b}. Additionally, \texttt{GRILLIX} is coupled with a streamlined diffusive neutral gas model that accounts for ionization, recombination, and charge-exchange reactions \cite{Eder2025, Horsten2017}. The fluid model assumes the plasma is highly collisional, such that the underlying distribution function closely approximates an isotropic Maxwellian. As a result, the model does not account for effects associated with non-thermal particles in specific regions of velocity space, such as trapped electrons.

Numerically, \texttt{GRILLIX} is based on the FCI approach and shares a significant portion of its code base with \texttt{GENE-X}. Specifically, the spatial mesh generation, equilibrium processing, and elliptic solver for the electromagnetic and electrostatic potentials are shared. The algorithm used in the elliptic solver is the same as described in the previous section. \texttt{GRILLIX} employs the finite difference method, with parallel operators discretized mimetically to reduce numerical perpendicular diffusion \cite{Stegmeir2016, Stegmeir2019}. At the target plates, sheath boundary conditions are implemented using the immersed boundary technique \cite{Stegmeir2023}. For time advancement, a third-order Karniadakis scheme \cite{Karniadakis1991} is applied, with the parallel heat flux terms treated implicitly.

\subsection{Local fluxtube delta-$f$ gyrokinetic simulations}\label{sec:gene_model}

For the present study, local fluxtube simulations were performed using
the gyrokinetic turbulence code \texttt{GENE} \cite{Jenko2000}. The code solves the delta-$f$ gyrokinetic equations including the full electromagnetic field fluctuations given in Ref. \cite{Pueschel2011}. The dynamics of two particle species (Deuterium and electrons) at real mass ratio are retained in this work. Collisions are included by means of a linear Sugama collision operator with appropriate conservation terms \cite{Crandall2020}. Effects of equilibrium rotation are neglected for the simulations
presented here.

The local version of \texttt{GENE} utilizes a field-aligned coordinate system $(x, y, z)$, where $x$, $y$, and $z$ represent the radial, binormal, and parallel directions with respect to the background magnetic field \cite{Beer1995}. The simulation domain is a magnetic flux tube, an elongated logically rectangular box that follows the magnetic field line. In the radial and binormal direction, a pseudo-spectral Fourier method is utilized, thereby retaining FLR effects to all orders through the use of Bessel functions. The flux surface geometry is extracted from LIUQE equilibria, in the form of G-EQDSK files, via a field line tracing method developed in Refs.~\cite{Xanthopoulos2006,Jenko2009}.

The gyrokinetic model in \texttt{GENE} incorporates a more comprehensive set of physics effects compared to the model used in \texttt{GENE-X}. Notably, higher-order FLR effects are included, parallel magnetic fluctuations are present, and advanced collision models are employed. In contrast, the use of field-aligned coordinates limits the geometric capabilities, allowing only closed flux surfaces to be simulated. Furthermore, the local fluxtube version restricts the simulation to local dynamics. The delta-$f$ approach requires fixed background profiles, enabling the simulation of fluctuations on top of these profiles while neglecting the self-consistent evolution of the background. The turbulence is driven by the local gradients of the background profiles. In \texttt{GENE-X}, both turbulence and profiles can evolve freely, with fluxes through the core boundary providing heat and particle sources.

\section{Simulation setup}

\begin{figure}
    \centering
    \includegraphics[width=0.75\textwidth]{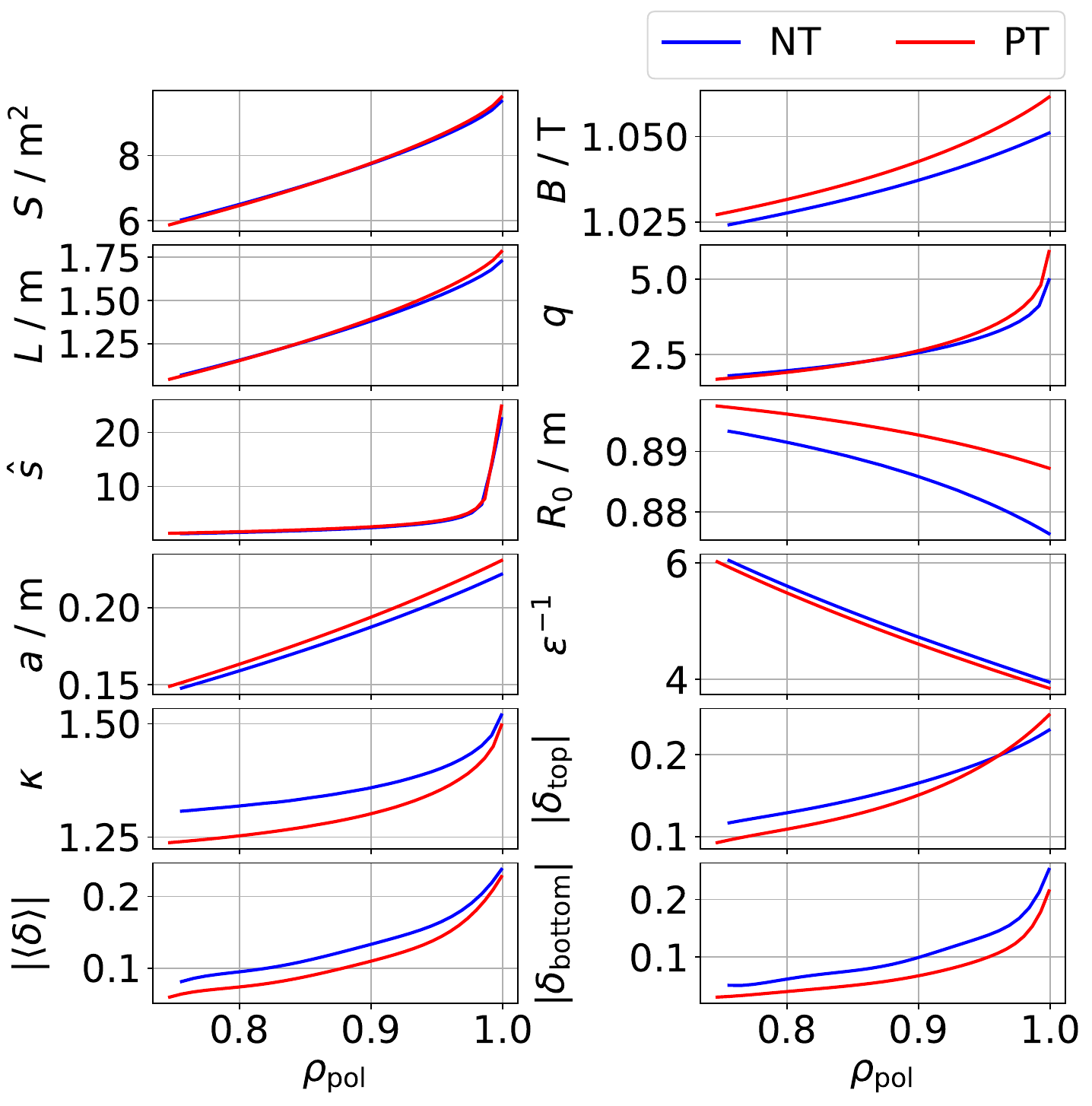}
    \caption{Profiles of various geometric quantities in NT and PT equilibria. The quantities shown include flux surface area $S$, flux surface arc length $L$ magnetic field absolute $B$, safety factor $q$, and normalized shear $\hat{s}=\partial \ln(q) / \partial \ln(r)$. The other quantities are the same as described in table \ref{tab:geom}.}
    \label{fig:geometry}
\end{figure}

We consider a pair of NT/PT magnetic equilibria from TCV, based on shots 68783, $t=1.6$ s (PT) and 68954, $t=1.0$ s (NT). Both cases are geometrically similar, except for the sign of the triangularity shaping parameter. Both cases are
attached L-mode discharges with an on-axis magnetic field of $B_0\approx 1.42$ T. The shaping parameters, as given in Ref. \cite{Sauter2016}, are listed in Table \ref{tab:geom}. Profiles of the geometric quantities are shown in Figure \ref{fig:geometry}. Notably, only minor differences in shaping parameters, except for the sign of $\delta$, are present in the given equilibria. It is worth noting that the triangularity shows a rather flat profile in these cases, with only low values of triangularity $\delta \lesssim0.1$ reached inside $\rho_\mathrm{pol}\sim0.9$. This results in flux surface shapes that are comparable in NT and PT. Further outside $\rho_\mathrm{pol}\gtrsim0.95$, shaping becomes more pronounced. Here $\rho_\mathrm{pol}=\sqrt{(\psi - \psi_\mathrm{axis})/(\psi_\mathrm{sep} - \psi_\mathrm{axis})}$ is the normalized poloidal flux surface label, where $\psi$ is the poloidal flux.

\begin{table}[]
\centering
\begin{tabular}{|l|c|c|c|c|c|c|c|}
\hline
   & $R_0$ / m & $a$ / m & $\epsilon$ & $\kappa$ & $\delta_\mathrm{top}$ & $\delta_\mathrm{bottom}$ & $\delta$ \\ \hline
NT & 0.876 & 0.222 & 0.253 & 1.554 & -0.230  & -0.252 & -0.238   \\ \hline
PT & 0.887 & 0.231 & 0.260 & 1.527 & 0.248   & 0.215  & 0.228    \\ \hline
\end{tabular}
\caption{Shaping parameters of the NT and PT equilibria calculated on the separatrix using equations (3) in Ref. \cite{Sauter2016}. The columns represent the major radius, minor radius, inverse aspect ratio, elongation, and top, bottom, and average triangularity, respectively.}
\label{tab:geom}
\end{table}

\begin{figure}
    \centering
    \includegraphics[width=0.5\textwidth]{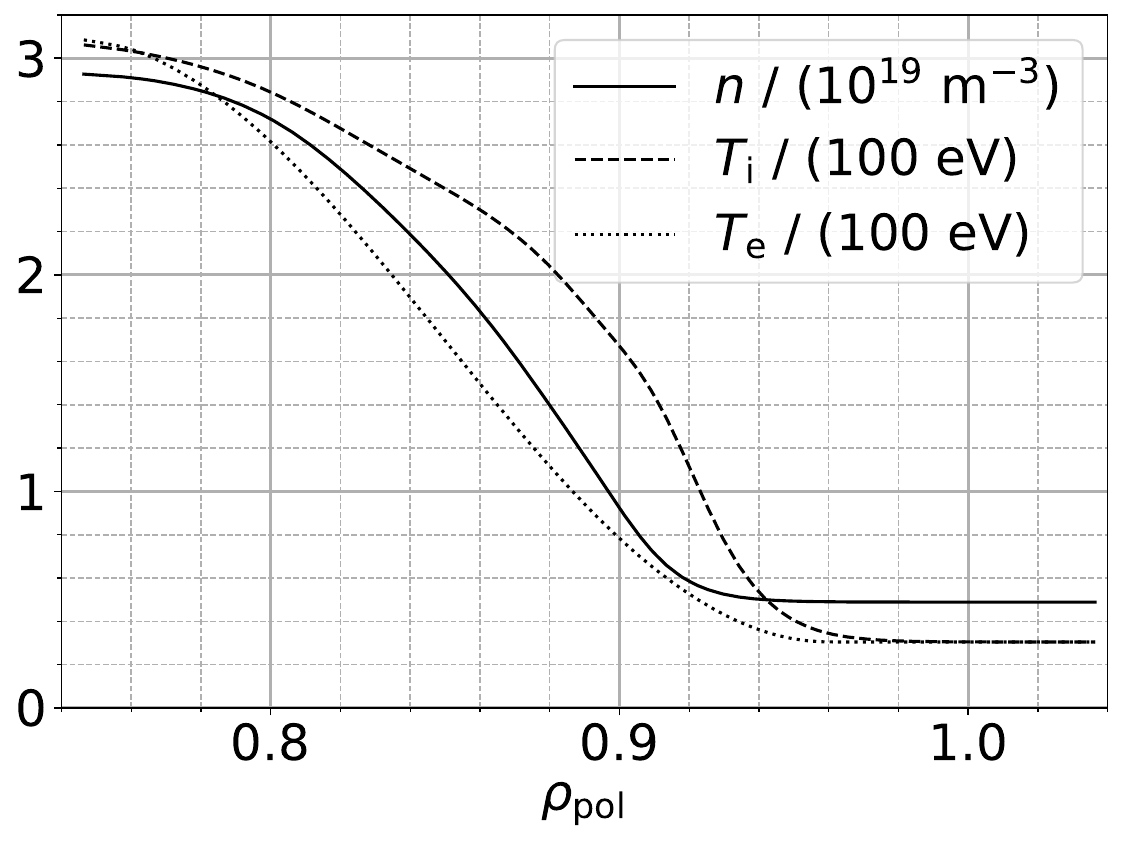}
    \caption{Initial profiles used to start the \texttt{GENE-X} simulation. The profiles show density (same for ions and electrons), ion and electron temperatures against the poloidal flux surface label $\rho_\mathrm{pol}$. The procedure to obtain these profiles is described in Ref. \cite{Ulbl2023}.}
    \label{fig:prof_initial}
\end{figure}

In this work, we conduct a numerical experiment where the plasma profiles, particularly the electron temperature, are increased at the core boundary of our simulations compared to the experimental values. As a result, the global density and temperature gradients are stronger than those observed in the experiment. In this context, \enquote{global} refers to the gradient from the inner to the outer simulation boundary, where profile values are fixed. Additionally, we expect a higher power flux to enter the simulation through the core boundary. The increase in electron temperature will lead to enhanced trapped electron mode drive of the developed turbulence, as shown later. The initial profiles are depicted in Figure \ref{fig:prof_initial}.

\subsection{\texttt{GENE-X} simulation parameters}

The equations are normalized to $L_\mathrm{ref}=0.903$~m, $B_\mathrm{ref} = 1.425$~T, $T_\mathrm{ref}=60$~eV and $n_\mathrm{ref}=10^{19}$~m$^{-3}$ (for details see Ref. \cite{Michels2021}). The resolution in the RZ plane is $\Delta RZ =$~1.228~mm resulting in approximately 177k points per plane, bounded by the $\rho_\mathrm{pol} \in [0.74, 1.04]$ flux surfaces. This corresponds to around 1.26 local Larmor radii $\rho_\mathrm{i}$ at the OMP separatrix with 1.15~T of magnetic field and initial ion temperature of 30~eV, and around 0.56$\rho_\mathrm{i}$ for the final ion temperature in the simulation of around 150~eV. A total of $N_\varphi=$~32 poloidal planes were used, and the velocity space is discretized by $N_{v_{||}}=$~80 and $N_{\mu}=$~32, spanning a box of $v_{||}\in [-8, 8] \times v_\mathrm{th}$ thermal velocities and $\mu \in [0, 81] \times T_\mathrm{ref}/B_\mathrm{ref}$. Two species, electrons and Deuterium ions with real mass ratio, were simulated. The time steps used for both cases are $(\Delta t)^\mathrm{PT} = 5.36$~ns and $(\Delta t)^\mathrm{NT} = 5.96$~ns ($0.00045$ and $0.0005$ in normalized units). Other parameters are the same as in Ref. \cite{Ulbl2023}.

In total, the distribution function has approximately 29 billion points, distributed across 256 MPI processes on 128 nodes of the A3 partition (Intel SkyLake) of the Marconi supercomputer at Cineca. The total computational cost per simulation is approximately 2-3 MCPUh on that machine.

\subsection{\texttt{GRILLIX} simulations parameters}

\texttt{GRILLIX} uses the same mesh as \texttt{GENE-X} but with a resolution of $\Delta RZ=$~0.99~mm, and $N_\varphi = 16$ poloidal planes resulting in approximately 260k points per plane. The timestep is $\Delta t=$1.3~ns in both cases. Free input parameters of the model are as follows. The heat flux limiting coefficients are set to $f^\mathrm{FS}_\mathrm{e} = 0.1$ and $f^\mathrm{FS}_\mathrm{i}=1.0$. The neutrals density at the target plates is fixed to $N_\mathrm{div}=5\cdot10^{16}$~m$^{-3}$. The simulation is initialized with sigmoid profiles, where the electron density is $n_\mathrm{e}= 2.5\cdot10^{19}$~m$^{-3}$ and the electron and ion temperatures are $T_\mathrm{e}=T_\mathrm{i}=280$~eV at the core boundary, $\rho_\mathrm{pol}=0.74$. The profiles decay to $n_\mathrm{e}= 5\cdot10^{18}$~m$^{-3}$ and $T_\mathrm{e}=T_\mathrm{i}=30$~eV in the SOL. The initial profile values near the core boundary are maintained throughout the simulation via an adaptive source, while the profile evolves freely and self-consistently elsewhere.

The \texttt{GRILLIX} simulations were performed on the Marconi A3 partition, with one MPI process assigned to each of the 16 poloidal planes, and 12 OpenMP threads allocated per plane. The simulations were run up to $1.6$~ms of simulated physics consuming approximately 55 kCPUh each.

\subsection{\texttt{GENE} simulations parameters}\label{sec:gene_sim}

For our linear simulations, we employed a default resolution of $(16\times32\times32\times18)$ points in the radial ($k_{x}$), parallel ($z)$, parallel velocity ($v_{||}$), and magnetic moment ($\mu$) grids, respectively. For the most challenging cases near the edge at $\rho_\mathrm{tor}=0.95$, a parallel ($z$) resolution of 64 points was used. The numerical convergence of our results was thoroughly verified in all dimensions. The radial box size was determined according to the standard setting of linear \texttt{GENE}, which uses the minimal box size allowed by the magnetic shear and chosen toroidal mode number to fit the parallel boundary condition.

In all simulations, the extent of the simulation domain along the magnetic field corresponds to one poloidal turn, and the velocity grids were set to encompass a region up to $|v_{\mathrm{max}}|=3v_{\mathrm{th},\alpha}$, with the thermal velocity $v_{\mathrm{th},\alpha}=\sqrt{2T_{0\alpha}/m_{\alpha}}$ of species $\alpha$. The simulations are performed using profiles obtained from the turbulent steady state in either the \texttt{GENE-X} or \texttt{GRILLIX} simulations. To ensure smooth profile gradients, time, toroidal, and poloidal averages are taken before using the resulting profiles in \texttt{GENE} simulations.

The linear simulations were carried out on the Cobra and Raven supercomputers at the Max-Planck Computing and Data Facility (MPCDF). The computational costs are negligible compared to the global gyrokinetic and fluid simulations.

\section{Simulation analysis}

\begin{figure}
    \centering
    \includegraphics[width=0.8\textwidth]{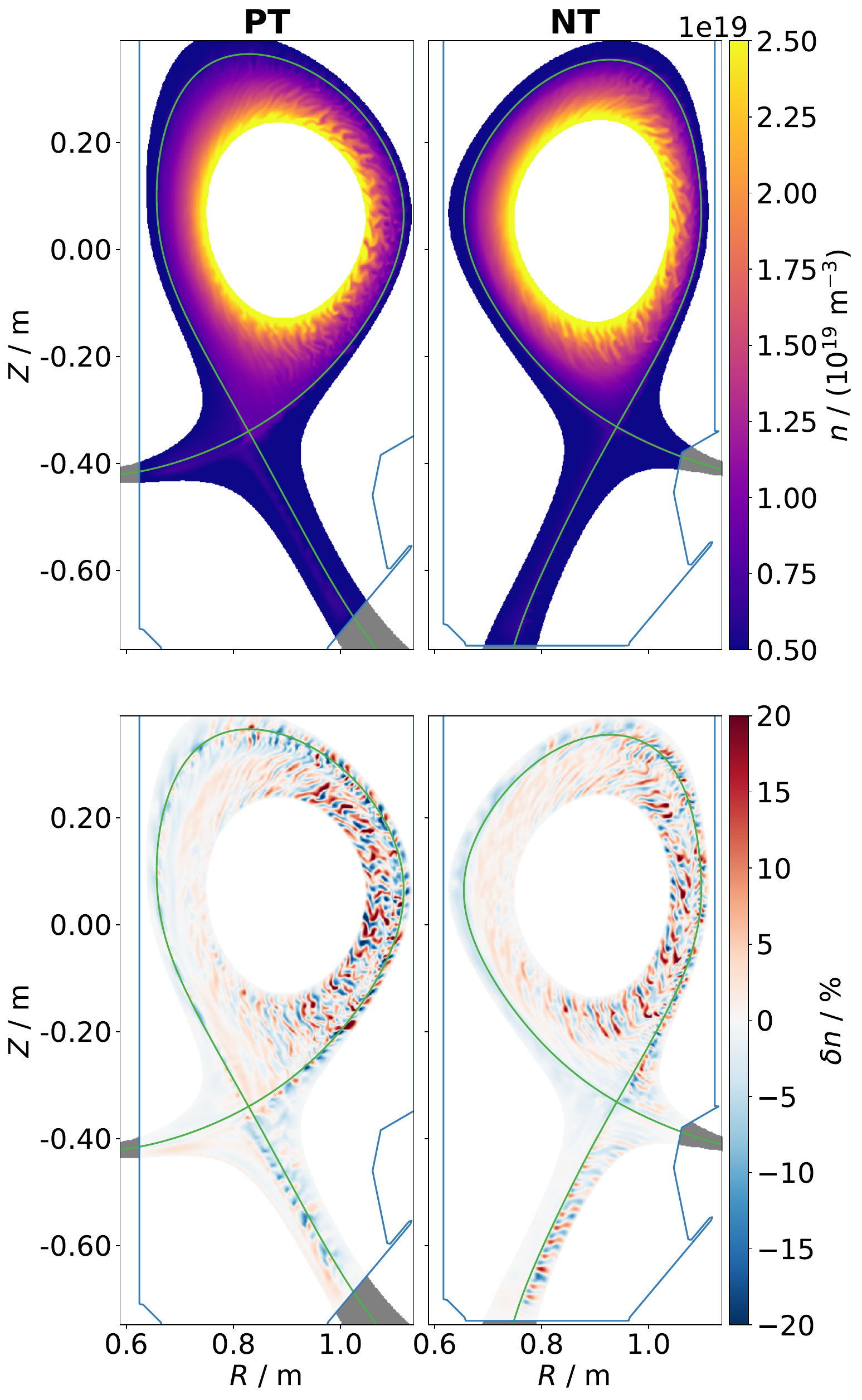}
    \caption{Snapshot of the final state in the \texttt{GENE-X} simulations in PT and NT. The total density is shown at a single poloidal plane, along with the fluctuations calculated as described in the main text. The temporal average has been applied over 100~$\mu$s (see figure \ref{fig:time_traces}) and the toroidal average over all planes in the simulation. The gray areas represent the divertor targets, where boundary conditions are enforced. The blue line shows the wall shape for reference (not used in the simulations), and the green line indicates the separatrix.}
    \label{fig:dens_snapshot}
\end{figure}

\begin{figure}
    \centering
    \includegraphics[width=0.65\textwidth]{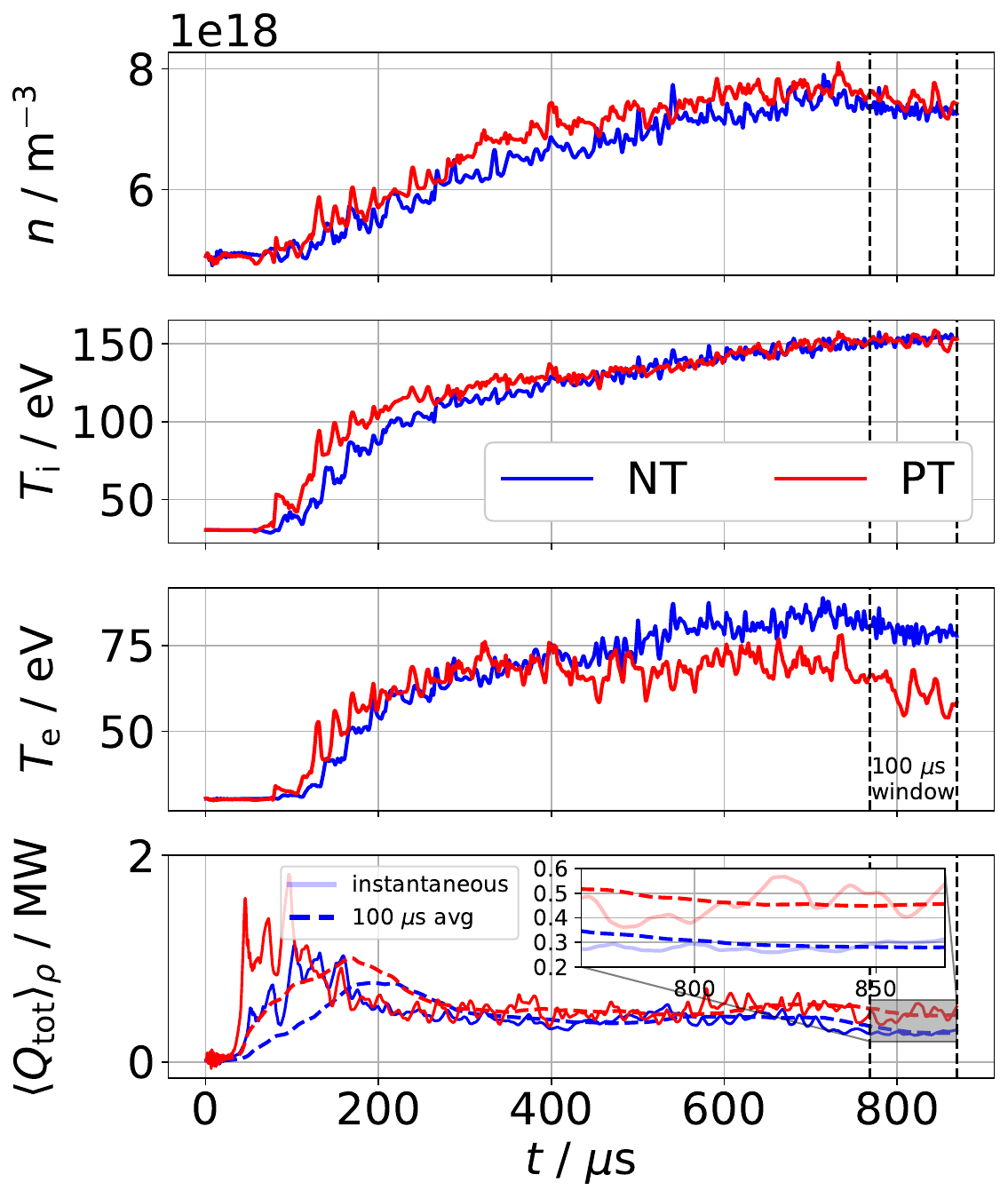}
    \caption{Time traces of density, ion and electron temperatures at a single point at the OMP separatrix and the total (electron + ion) radially averaged heat flux in the \texttt{GENE-X} simulations. The black dashed lines indicate the 100~$\mu$s time window used for providing temporally averaged analysis data. The heat flux is given as the sum of electrons and ions and ExB and diamagnetic contributions from eqs. (\ref{eq:exb_heatflux}-\ref{eq:dia_heatflux}). The heat flux is provided as instantaneous time traces (solid lines) as well as temporally rolling averaged (dashed).}
    \label{fig:time_traces}
\end{figure}

Global turbulence simulations were performed from the initial condition (figure~\ref{fig:prof_initial}) until the developed turbulence reached a quasi-stationary state. An example for this turbulent state is given by the final density $n$ snapshots and the corresponding fluctuations $\delta n = (n - \left<n\right>_{t,\varphi}) / \left<n\right>_{t,\varphi}$ shown in figure~\ref{fig:dens_snapshot}. The bracket $\left<\phantom{n}\right>_{t,\varphi}$ denotes a temporal and toroidal average operation. Time traces from the initial simulation start through the onset of turbulence up to the approximately saturated phase are given in figure~\ref{fig:time_traces}.

As a general result, the turbulence developed in the \texttt{GENE-X} simulation of the PT case is significantly stronger than in NT. This is exemplified in figure \ref{fig:dens_snapshot} and applies to temperature and electrostatic potential fluctuations as well. Consequently, the electrostatic ExB heat flux is reduced in NT, and the electron temperature is somewhat larger than in PT. Notably, this trend is not necessarily present throughout the whole simulation time but rather develops within the nonlinear phase and turbulence saturation.

In the following, a comprehensive analysis of the \texttt{GENE-X} and \texttt{GRILLIX} simulations is conducted, and the \texttt{GENE} code is utilized to provide further physics insights into the observed turbulence.

\subsection{Profiles and heat fluxes}

\begin{figure}
    \centering
    \includegraphics[width=0.65\textwidth]{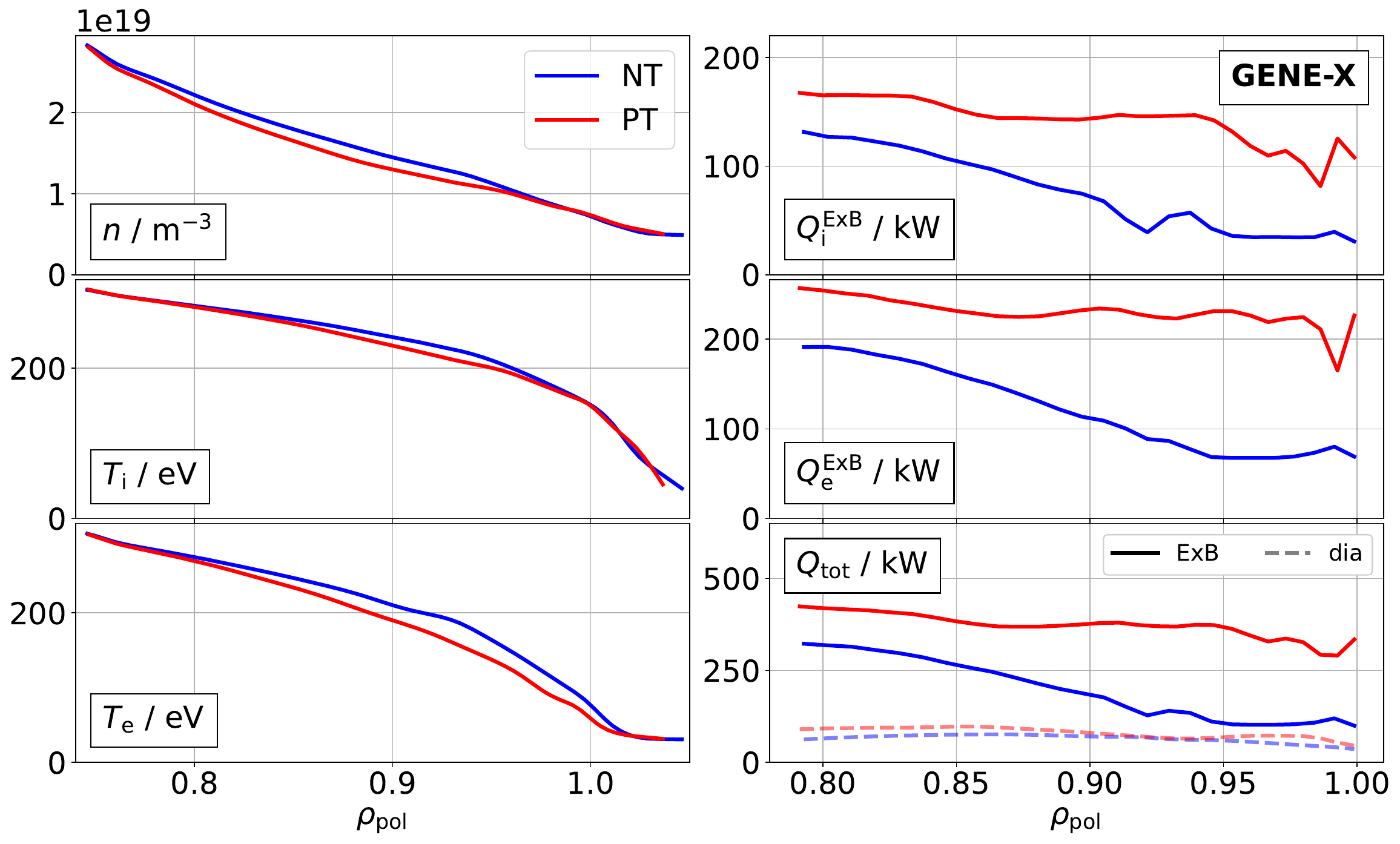}
    \caption{OMP profiles of temporally and toroidally averaged density, ion and electron temperatures, and flux-surface integrated heat fluxes in the \texttt{GENE-X} simulations. For the heat fluxes the ExB components are shown and additionally the diamagnetic contributions for the total (electron + ion) heat flux. The heat fluxes were calculated using (\ref{eq:exb_heatflux}) and (\ref{eq:dia_heatflux}) on flux surfaces considering only the confined region.}
    \label{fig:profiles_genex}
\end{figure}

\begin{figure}
    \centering
    \includegraphics[width=0.65\textwidth]{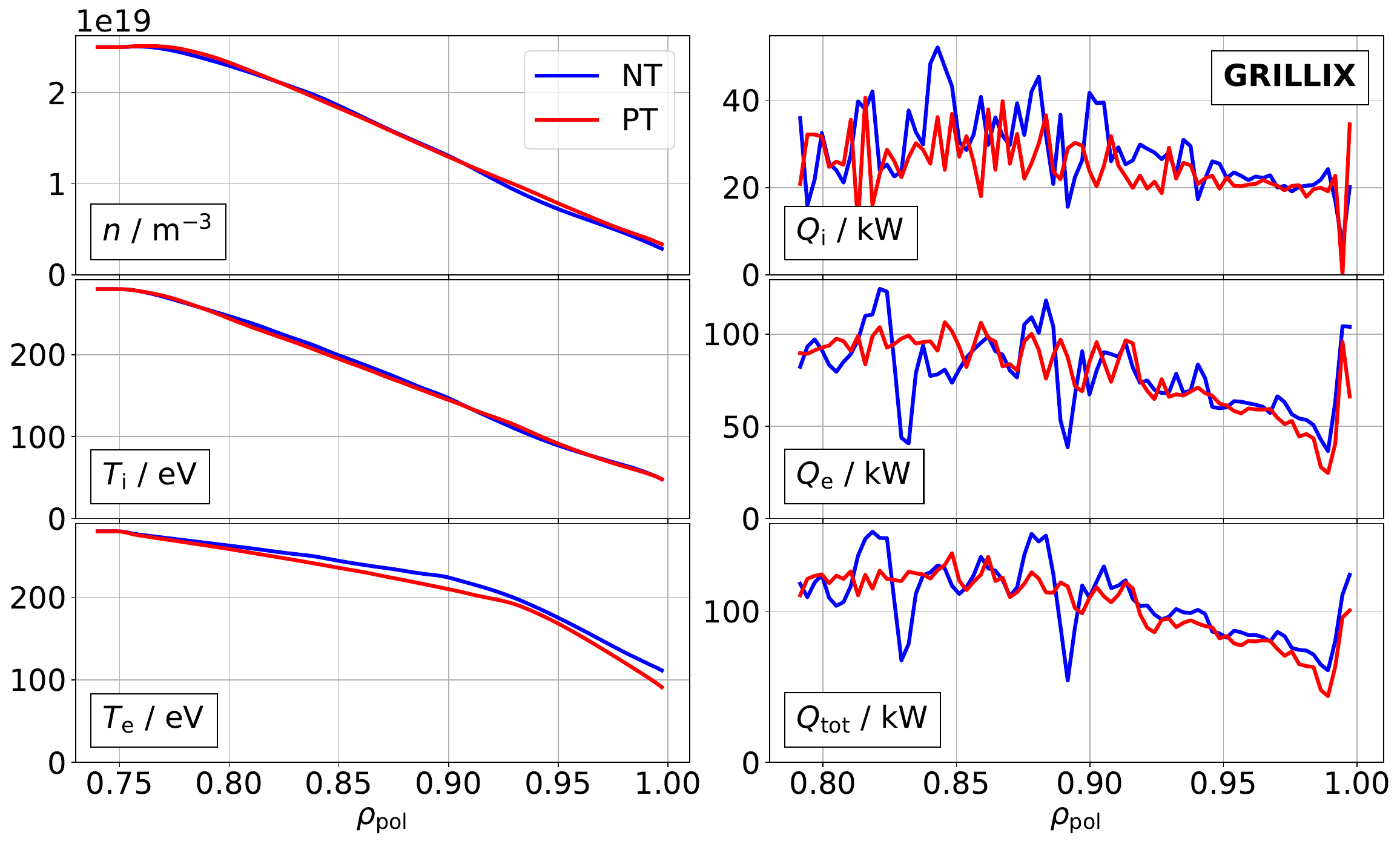}
    \caption{OMP profiles and integrated heat fluxes for the \texttt{GRILLIX} simulation, otherwise the same as figure \ref{fig:profiles_genex}.}
    \label{fig:profiles_grillix}
\end{figure}

Plasma profiles measured at the OMP are presented in figure \ref{fig:profiles_genex} for the gyrokinetic \texttt{GENE-X} simulations and in figure \ref{fig:profiles_grillix} for drift-fluid \texttt{GRILLIX} simulations. Furthermore, total radial heat flux is given, approximated by the ExB and diamagnetic heat fluxes. Electromagnetic heat fluxes due to magnetic flutter are comparatively small and neglected. The heat fluxes follow from the gyrokinetic model (\ref{eq:gyrokin}) as

\begin{align}
    Q_\alpha^\mathrm{ExB} &\approx \oiint_\rho \frac{c}{B}\left\{\mathbf{b}\times\nabla\phi_1\right\}_\rho W_\alpha R(l) \mathrm{d}l \mathrm{d}\varphi, \label{eq:exb_heatflux},\\
    Q_\alpha^\mathrm{dia} &\approx \oiint_\rho \frac{c}{q_\alpha n_\alpha B} \Bigg[\frac{1}{B}\left\{\mathbf{b}\times\nabla B\right\}_\rho \big(W_{\perp,\alpha}W_{||,\alpha} + 2 W_{\perp,\alpha}^2 \big) \nonumber\\
    &\hspace{2.8cm} + \left\{\nabla \times \mathbf{b}\right\}_\rho \big(2W_{\perp,\alpha}W_{||,\alpha} + 6 W_{||,\alpha}^2 \big)
    \Bigg] R(l) \mathrm{d}l \mathrm{d}\varphi.\label{eq:dia_heatflux}
\end{align}

Here the closed double integral denotes integration over a full flux surface (for fixed $\rho$), where $\{\phantom{n}\}_\rho$ represents the radial projection and $l$ denotes the poloidal flux surface arc length. The energy moment is denoted by $W_\alpha = 3 T_\alpha / (2 n_\alpha)$. The parallel and perpendicular energies are defined as $W_{||,\alpha} = T_{||,\alpha} / (2 n_\alpha)$ and $W_{\perp,\alpha} = T_{\perp,\alpha} / n_\alpha$, where only the corresponding parallel/perpendicular terms in the velocity space moments (\ref{eq:T_mom}) are considered.

The diamagnetic heat flux (\ref{eq:dia_heatflux}) includes only the equilibrium contribution from an anisotropic Maxwellian. The term \enquote{diamagnetic heat flux} encompasses contributions from both grad-B and curvature drifts to the heat flux, as represented by the first and second terms in the square bracket. In the limit of an isotropic Maxwellian, where $T_{||,\alpha},T_{\perp,\alpha}\to T_\alpha$, standard formulas for the diamagnetic heat flux are obtained, which are  $Q_\alpha^\mathrm{dia}\propto 5 n_\alpha T_\alpha / 2$ (see e.g.~equation~(9) in Ref.~\cite{Zholobenko2024}). The contributions of the diamagnetic heat fluxes are noticeable compared to the ExB heat fluxes in this case, and amount to approximately 100 kW.

The radial heat fluxes in \texttt{GRILLIX} are computed according to,
\begin{align}
    Q_\alpha^\mathrm{GRILLIX} =  \oiint_\rho \frac{c}{B} \left[\frac{3}{2}p_\alpha\big\{\mathbf{b}\times\nabla\phi_1\big\}_\rho    + \frac{5}{2}\frac{1}{q_\alpha }\big\{\mathbf{b}\times\nabla(p_\alpha T_\alpha)\big\}_\rho\right]\, \mathrm{d}l \mathrm{d}\varphi \label{eq:exb_heatflux_grillix},
\end{align}
where the two terms denote ExB and diamagnetic heat flux respectively. Electromagnetic heat fluxes were found to be negligible in the simulations. Here, $p_\alpha$ and $T_\alpha$ denote plasma pressure and temperature in the fluid model.


The plasma profiles (figures~\ref{fig:profiles_genex},~\ref{fig:profiles_grillix}) developed in the stationary turbulent state are remarkably similar in NT and PT, both in gyrokinetic and drift-fluid simulations. However, the situation is different for the heat fluxes. Our gyrokinetic simulations show substantially different heat fluxes between NT and PT. The 100~$\mu$s averaged heat fluxes at the end of the simulations (from figure~\ref{fig:time_traces}) are $Q_\mathrm{tot}^\mathrm{NT}\approx298$~kW and around 57\% more in PT, namely $Q_\mathrm{tot}^\mathrm{PT}\approx468$~kW. The primary difference between NT and PT is observed in the turbulent ExB heat fluxes. The diamagnetic contribution, while contributing noticeably, is about the same in NT and PT. As a remark regarding the NT simulation, the heat flux profile indicates that the simulation is close to saturation, but not fully saturated. The heat fluxes decrease radially, leading to profile steepening. These moderate dynamics do not affect the qualitative result presented here. In contrast to the gyrokinetic results, the drift-fluid simulations exhibit almost identical and significantly smaller heat fluxes in NT and PT.

The results obtained in the gyrokinetic simulation are qualitatively consistent to experimental findings in Ref. \cite{Camenen2007}. In that study, matching NT and PT discharges in TCV were conducted resulting in comparable profiles but reduced heat flux in NT, where only half the heating power was required. Turbulence in those discharges was found to be dominated by the trapped electron mode (TEM) \cite{Kadomtsev1971}. Subsequent linear and non-linear local gyrokinetic studies \cite{Marinoni2009} have identified the improved confinement in NT as a result of the stabilization of the underlying TEMs by a modification of the trapped particle toroidal precession drift. Indeed, as demonstrated by the detailed turbulence characterization presented later (section \ref{sec:turb}), the \texttt{GENE-X} simulations reveal that the turbulence is driven by the trapped electron mode (TEM). In contrast, the turbulence in the drift-fluid simulations is found to be rather drift-wave-like. These results can be attributed to the fact that the drift-reduced Braginskii codes do not account for the effects of trapped particles. The difference observed between the \texttt{GENE-X} and \texttt{GRILLIX} simulations is therefore presumably due to the presence of trapped particle physics. Moreover, the difference in heat flux between NT and PT can be attributed to the influence of triangularity on the TEM \cite{Marinoni2009}. This suggests that the improved confinement in NT is a direct result of the modification of the TEM by the triangularity.

\subsection{Energy confinement}

To quantify the improved energy confinement in NT, we calculate the energy confinement time, $\tau_E = W_\mathrm{tot}/Q_\mathrm{tot}$, where $W_\mathrm{tot}$ denotes the total stored energy and $Q_\mathrm{tot}$ the total heating power. The total stored energy is obtained by integrating the total pressure over the full flux surface volume
\begin{align}
    W_\mathrm{tot} = \iiint \frac{3}{2}\big( n_\mathrm{e} T_\mathrm{e} + n_\mathrm{i} T_\mathrm{i} \big) \frac{\partial r(\rho, \theta)}{\partial \rho}  r(\rho, \theta) R(\rho, \theta) \mathrm{d}\rho \mathrm{d}\theta \mathrm{d}\varphi,
\end{align}
in a toroidal $(\rho, \theta, \varphi)$ coordinate system. The total heating power is approximated by summing the contributions from ExB and diamagnetic heat fluxes, $Q_\mathrm{tot} = \sum_\alpha \left< Q_\alpha^\mathrm{ExB} + Q_\alpha^\mathrm{dia} \right>_{\rho}$, assuming a steady-state where the heat fluxes are approximately constant across flux surfaces.

\begin{figure}
    \centering
    \includegraphics[width=0.5\textwidth]{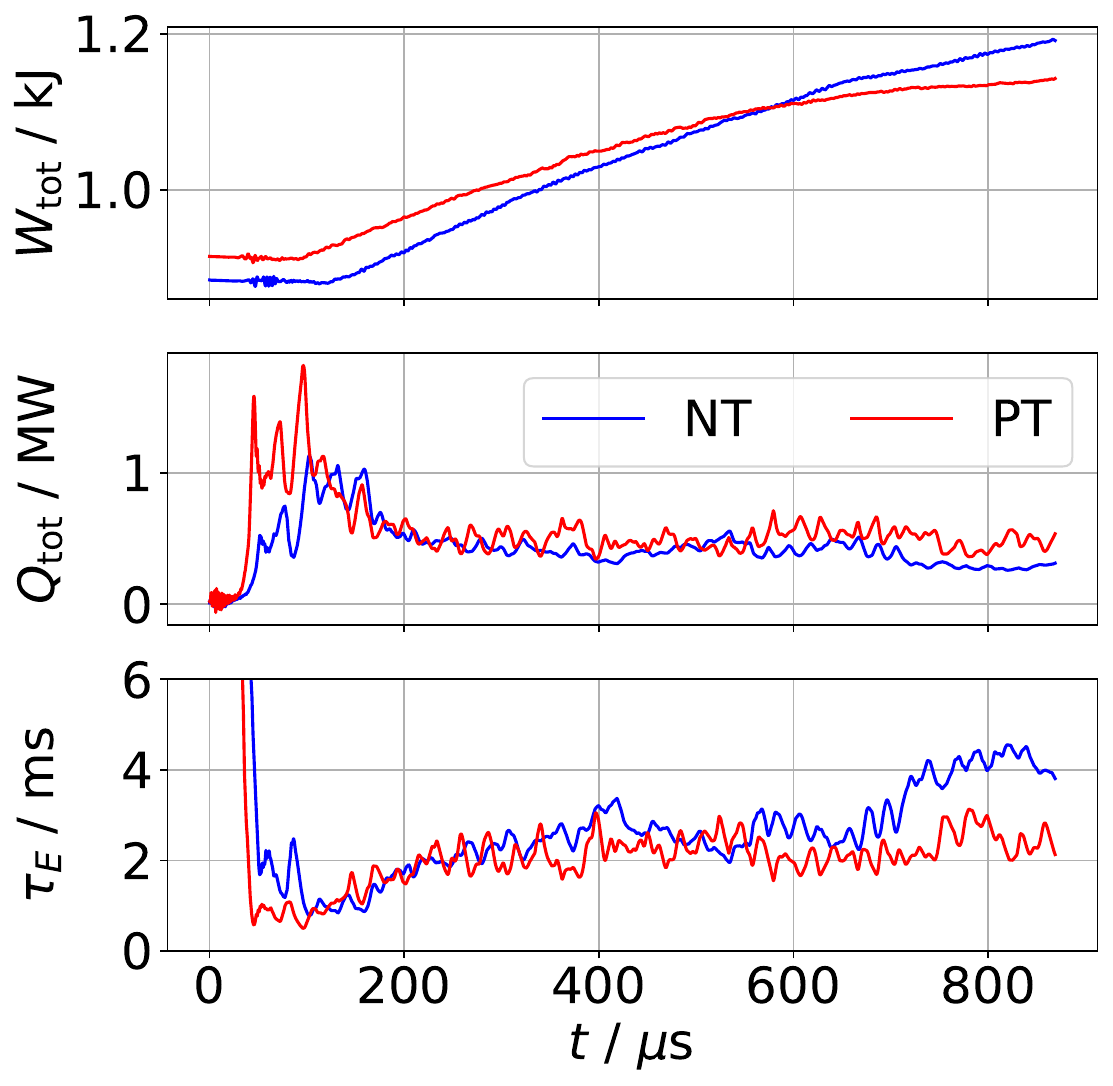}
    \caption{Time traces of total stored energy $W_\mathrm{tot}$, total heating power $Q_\mathrm{tot}$ and energy confinement time $\tau_E$ for NT and PT in the \texttt{GENE-X} simulations. 
    }
    \label{fig:energy_conft}
\end{figure}

The time traces of total stored energy, total heating power, and energy confinement time are shown in figure \ref{fig:energy_conft}. We note that the total stored energy has only been calculated within the closed field line region, in a poloidal annulus $\rho_\mathrm{pol}\in[0.74, 0.999]$. As a result, the confinement time in the plasma edge is significantly smaller compared to including the core. The initial state in PT contains more energy than NT, which is attributed to the difference in plasma volume, which is around 8\% smaller in NT. Furthermore, the different poloidal flux profile affects the initial energy content due to the identical profiles being initialized as a function of poloidal flux.
Regarding the temporal evolution, we observe that PT experiences an earlier energy build-up in the confined region. However, in the non-linear saturation phase, the stored energy in NT surpasses PT, resulting in an increase in confinement time. The 100~$\mu$s averaged confinement time at the end of the simulations is $\tau_E^\mathrm{NT}\approx 4.22$~ms and $\tau_E^\mathrm{PT}\approx 2.52$~ms. NT exhibits a 67\% increase of energy confinement in this case.

\subsection{Turbulence characterization}\label{sec:turb}

\begin{figure}
    \centering
    \includegraphics[width=0.9\textwidth]{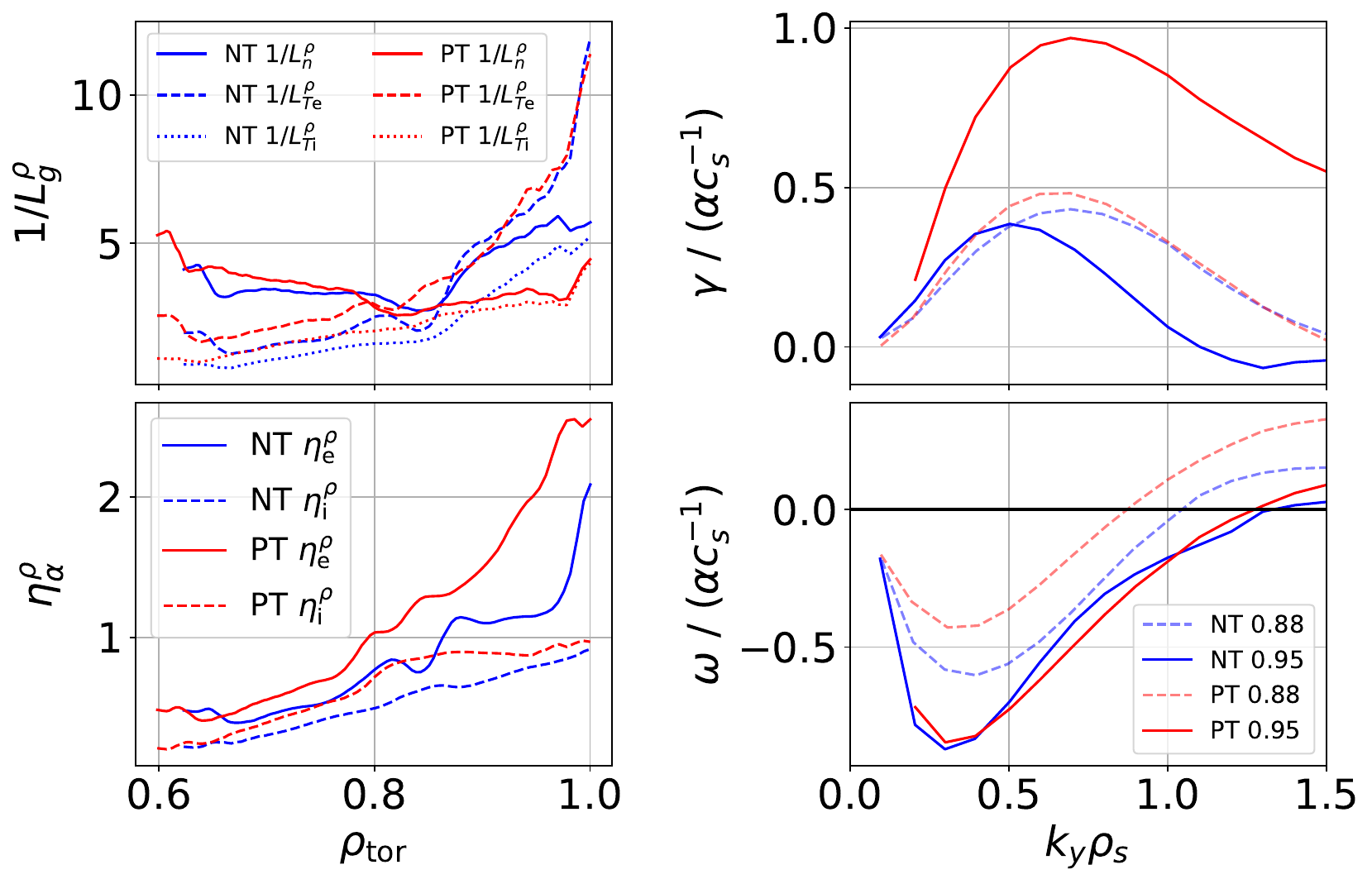}
    \caption{Linear gyrokinetic fluxtube simulations using \texttt{GENE} on averaged profiles of the saturated turbulent state in the \texttt{GENE-X} simulations in NT and PT. The left plots show the inverse gradient lengths $L_{T\mathrm{e}}^\rho=-\partial \ln(T_\mathrm{e})/\partial \rho_\mathrm{tor}$ and their ratios $\eta_\alpha^\rho = L_{n\alpha}^\rho/L_{T\alpha}^\rho$. The right plots show the resulting growth rates and frequencies. The binormal wavenumber is scaled by the local sound speed and an effective minor radius defined as $\alpha=\sqrt{\psi_\mathrm{tor}/(\pi B_{\mathrm{tor},0})}\sim a$, where $\psi_\mathrm{tor}$ is the toroidal flux and $B_{\mathrm{tor},0}$ the on-axis toroidal magnetic field.
    }
    \label{fig:gene}
\end{figure}

To identify the expected dominant micro-instability in the non-linear, saturated turbulent state of the \texttt{GENE-X} simulation, we perform linear gyrokinetic fluxtube simulations using \texttt{GENE} to investigate the growth rates and frequencies as a function of binormal wavenumber $k_y$. To prepare the \texttt{GENE-X} simulation data for the linear gyrokinetic fluxtube simulations, we interpolate the electron and ion densities and temperatures onto a $(\rho_\mathrm{pol}, \theta) = (100, 30)$ polar grid. We then perform an average over angles $\theta$ and $\varphi$, as well as over 100~$\mu$s in time. The resulting averaged profiles are then converted to a toroidal flux surface label by solving the equation $q = \mathrm{d}\psi_\mathrm{tor}/\mathrm{d}\psi_\mathrm{pol}$ \cite{Dhaeseleer}, which yields $\psi_\mathrm{tor}(\rho_\mathrm{pol})=\int_0^{\rho_\mathrm{pol}} q(\rho_\mathrm{pol}' ) \rho_\mathrm{pol}' \mathrm{d}\rho_\mathrm{pol}'$ where constant prefactors can be neglected due to re-normalizing $\rho_\mathrm{tor}=\sqrt{(\psi_\mathrm{tor} - \psi_\mathrm{tor,axis})/(\psi_\mathrm{tor,sep} - \psi_\mathrm{tor,axis})}$. The local gradients lengths and their ratios are shown in figure~\ref{fig:gene}.

As a result, \texttt{GENE} predicts the trapped electron mode (TEM) to be the dominant micro-instability throughout the entire considered profile range $\rho_\mathrm{tor}\in[0.8,0.95]$. Figure~\ref{fig:gene} shows the growth rates and frequencies for two radial locations, exemplifying the predicted behavior. The growth rates peak at ion scales $k_\mathrm{y}\rho_s \lesssim 1$ and the frequencies are directed in the electron direction. The sound speed $c_s=\sqrt{T_\mathrm{e}/m_\mathrm{i}}$ and the sound Larmor radius $\rho_s=c\sqrt{T_\mathrm{e}m_\mathrm{i}}/(e B)$ are used to quantify the scales and amplitudes of the instabilities.

\begin{figure}
    \centering
    \includegraphics[width=0.9\textwidth]{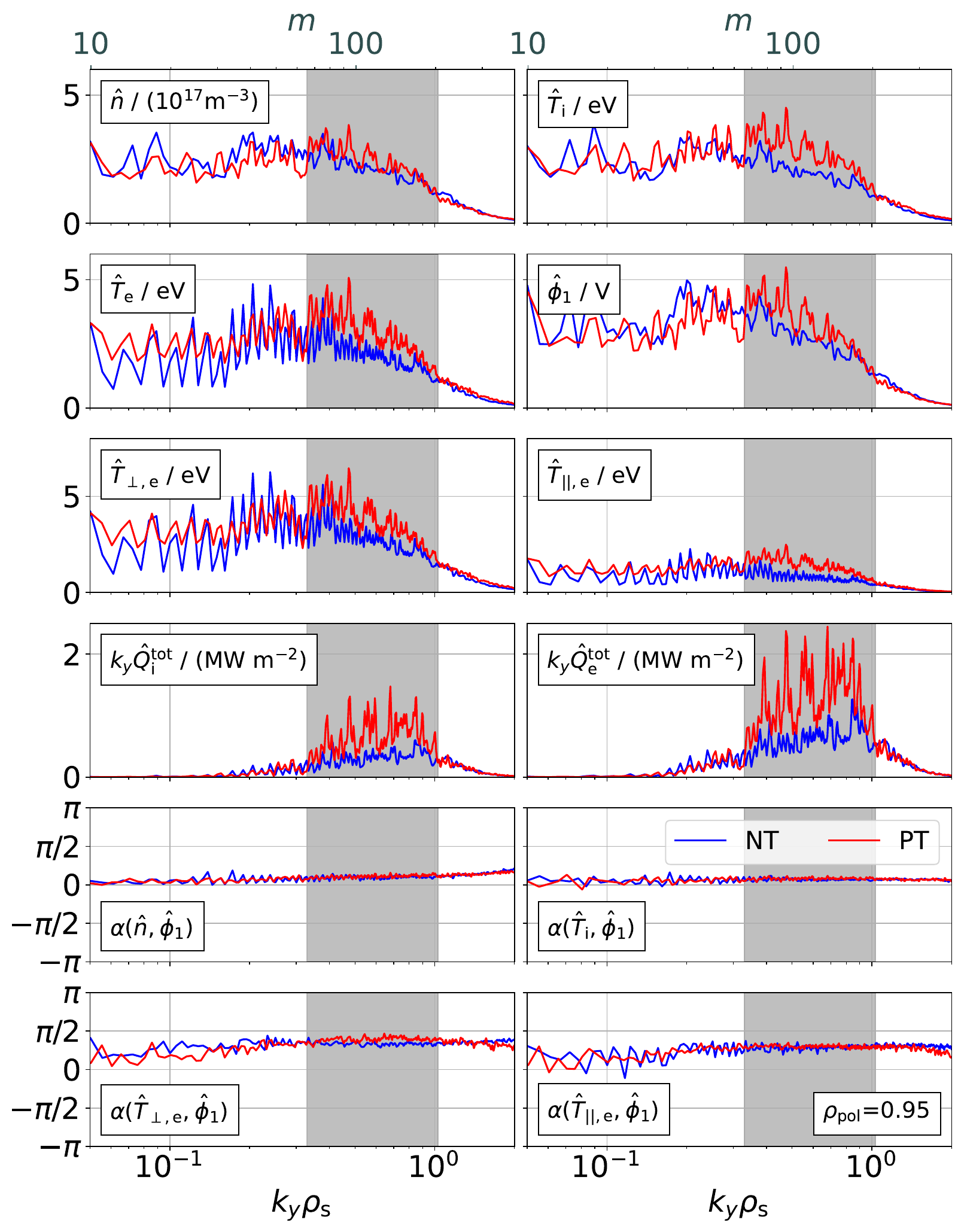}
    \caption{Summary of spectral turbulence characterization on the flux surface $\rho_\mathrm{pol}=0.95$ in the \texttt{GENE-X} simulations. Rows 1-3 display density, temperature, and electrostatic potential spectra. Row 4 shows spectral heat fluxes $k_y \hat{Q}^\mathrm{tot}_\mathrm{\alpha}$, where the area under the curves represents the total integral heat flux through the flux surface. Rows 5-6 show average phase shifts between density and temperatures and the electrostatic potential. The second x-axis on the top indicates the poloidal mode number $m$ based on the PT simulation ($m$ for NT is slightly different due to the different local temperatures). The gray shaded area highlights the approximate spectral region where the heat flux is large.}
    \label{fig:spectral_analysis}
\end{figure}

We compare the linear results by characterizing the turbulence in the non-linear global \texttt{GENE-X} simulations using previously developed Fourier diagnostics \cite{Ulbl2023} on the $\rho_\mathrm{pol}=0.95$ ($\rho_\mathrm{tor}\approx0.88$) flux surface. We use slightly modified definitions from Ref.~\cite{UlblPhd}. Figure~\ref{fig:spectral_analysis} summarizes the relevant findings. Note that in \texttt{GENE-X}, $k_y$ represents the poloidal wavenumber, whereas in \texttt{GENE}, $k_y$ represents the binormal wavenumber. While the poloidal and binormal wavenumbers differ due to the field line tilt, they can still be used for qualitative comparisons. In this analysis, we use flux surface-averaged quantities, such as the temperature, to evaluate the local sound Larmor radius. We also evaluate the spectral ExB heat fluxes using the method described in Ref. \cite{Ulbl2023, Zholobenko2023},
\begin{align}
    \hat{Q}_\alpha^\mathrm{conv} &= \frac{3 r \left<T_\alpha\right>_y}{2\left<B\right>_y} k_y |\hat{n}_\alpha| \, |\hat{\phi}_1| \, \sin\Big(\alpha\big(\hat{n}_\alpha, \hat{\phi}_1\big)\Big),\label{eq:spec_conv}\\
    \hat{Q}_\alpha^\mathrm{cond} &= \frac{3 r \left<n_\alpha\right>_y}{2\left<B\right>_y} k_y |\hat{T}_\alpha| \, |\hat{\phi}_1| \, \sin\Big(\alpha\big(\hat{T}_\alpha, \hat{\phi}_1\big)\Big),\label{eq:spec_cond}
\end{align}
which represent convective and conductive contributions respectively. The total heat flux is given by $\hat{Q}_\alpha^\mathrm{tot}=\hat{Q}_\alpha^\mathrm{conv}+\hat{Q}_\alpha^\mathrm{cond}$. Here $r=L/(2\pi)$ represents the local effective minor radius, calculated from the flux surface arc length. The flux surface line average of a function $g$ is denoted by $\left<g\right>_y$. Fourier transformed quantities are denoted by $\hat{g}$. Phase shifts are calculated via $\alpha(g,h) = \mathrm{Im}\big(\mathrm{log}(g^* h)\big)$, where $g^*$ denotes the complex conjugate.  Additionally, we perform a temporal Fourier transform on the electrostatic potential to obtain a $\omega(k)$ dispersion relation of the various Fourier modes present \cite{Ulbl2023}.

\begin{figure}
    \centering
    \includegraphics[width=0.75\textwidth]{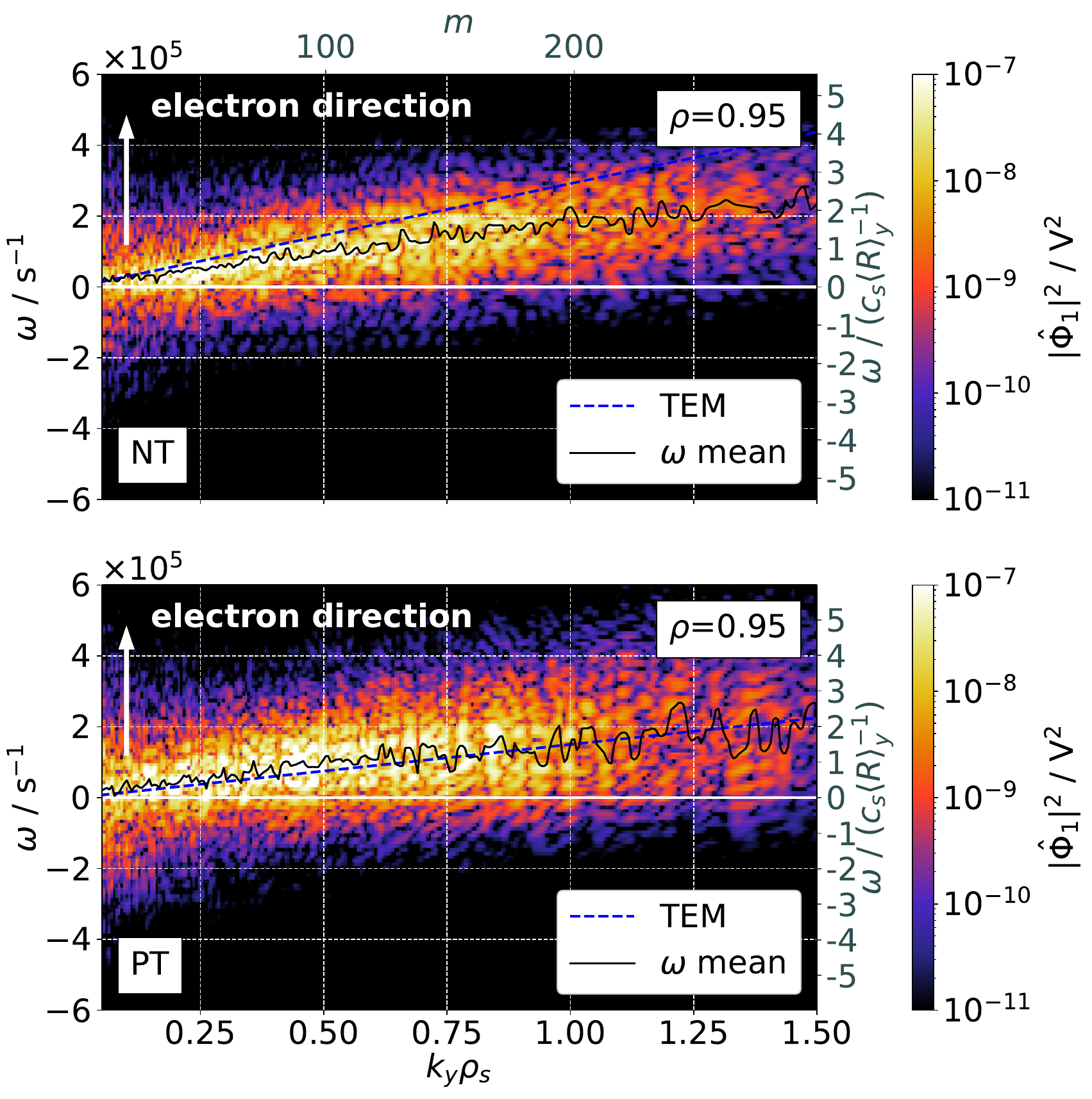}
    \caption{Dispersion relation on the flux surface $\rho_\mathrm{pol}=0.95$ in the \texttt{GENE-X} simulations. Shown are the Fourier amplitudes of the electrostatic potential $\hat{\Phi}_1$, which is the Fourier transform of $\hat{\phi}_1$ in time. The black line represents the average frequency of the underlying dispersion, while the blue dashed line shows the linear TEM frequency obtained from (\ref{eq:tem}).}
    \label{fig:dispersion}
\end{figure}

The spectral analysis of the \texttt{GENE-X} simulations reveals several indications of the presence of TEMs in the system. Firstly, the electrostatic electron ExB heat flux dominates, particularly its contribution from perpendicular temperature fluctuations. Secondly, the electron temperature phase shifts with respect to $\phi_1$ are between $\pi/4$ and $\pi/2$, indicating interchange drive. In contrast, the phase shifts of density and ion temperature fluctuations are more drift-wave-like, with values closer to 0. Thirdly, the dispersion relations (figure \ref{fig:dispersion}) confirm that the modes propagate primarily in the electron direction. Furthermore, the measured dispersion is well approximated by the linear dispersion relation of the collisionless TEM in the fluid limit \cite{Kadomtsev1967,BrunnerPhd,Garbet2024}, which is obtained by solving for the real part of the dispersion relation
\begin{align}
    \omega_\mathrm{TEM}^2 + \omega_\mathrm{TEM} \frac{\big(\omega_{n\mathrm{e}} - \frac{3}{2} \omega_{\varphi\mathrm{e}}\big) f_\mathrm{t} - \omega_{n\mathrm{e}}}{1-f_\mathrm{t}} + \frac{\frac{3}{2} \omega_{n\mathrm{e}} \omega_{\varphi\mathrm{e}} (1 + \eta_\mathrm{e}) f_\mathrm{t}}{1-f_\mathrm{t}} = 0,\label{eq:tem}
\end{align}
where $f_\mathrm{t}\sim\sqrt{1-B_\mathrm{min}/B_\mathrm{max}}$ denotes the trapped particle fraction, gradient length ratios $\eta_\alpha=L_{n\alpha}/L_{T\alpha}$ are defined using $L_{g}=-1/\nabla\ln(g)$, the diamagnetic drift frequency is given by $\omega_{n\alpha}=T_\alpha/(q_\mathrm{e} B L_{n\alpha}) k_y$, and the toroidal precession drift frequency is approximated by the curvature drift frequency, $\omega_{\varphi\alpha}\approx \omega_{n\alpha} L_{n\alpha}/R$.

Additionally, we performed linear \texttt{GENE} simulations at radial locations $\rho_\mathrm{tor}=0.88$ and $\rho_\mathrm{tor}=0.95$ of the saturated turbulent state in \texttt{GRILLIX}. The results suggest that TEMs are dominant in the local gyrokinetic fluxtube simulations. However, the drift-reduced Braginskii fluid model used in \texttt{GRILLIX} does not include trapped particle physics, which is a key aspect of the TEM. We characterized the turbulent state of the \texttt{GRILLIX} simulations using the same methods as for the \texttt{GENE-X} simulation above. The results show a different picture in \texttt{GRILLIX}, with drift-wave turbulence present in both NT and PT, and some ITG contributions at scales $k_y\rho_s\approx0.25$ in PT only. This cross-verification highlights the importance of using a gyrokinetic model, particularly in cases like this where the presence of TEMs is crucial. Furthermore, the beneficial effect of NT is clearly connected to the presence of TEMs in the simulations.

\begin{figure}
    \centering
    \includegraphics[width=0.9\textwidth]{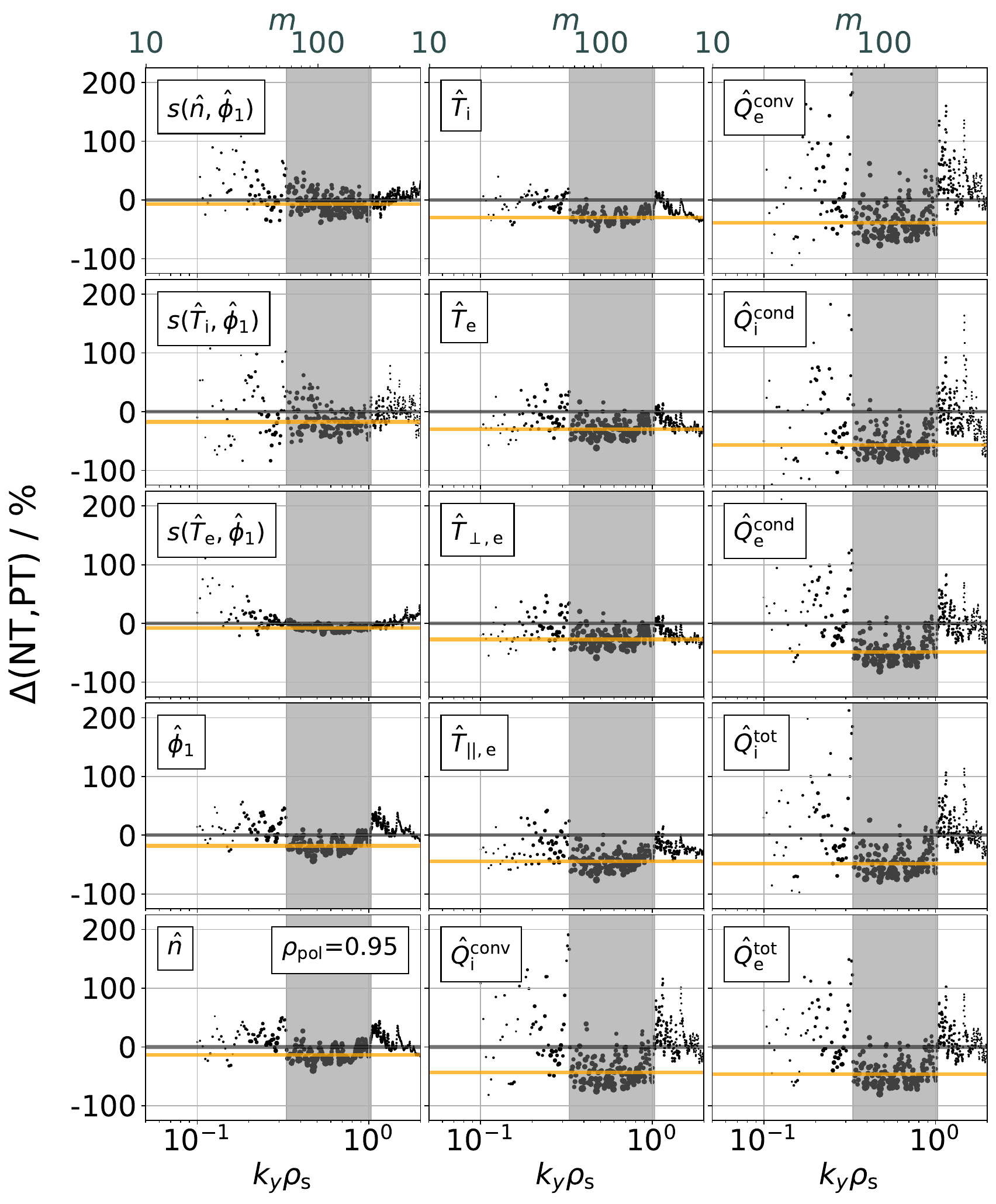}
    \caption{Comparison of individual terms that compose the spectral heat fluxes (\ref{eq:spec_conv}-\ref{eq:spec_cond}) in NT and PT \texttt{GENE-X} simulations. The abbreviation $s(\hat{n},\hat{\phi}_1)\equiv \mathrm{sin(\alpha(\hat{n},\hat{\phi}_1))}$ is used. The difference between NT and PT is calculated as $\Delta(\mathrm{PT},\mathrm{NT}) = (\mathrm{NT} - \mathrm{PT}) / \mathrm{PT}$ for each quantity, showing the increase in PT compared to NT. Positive values indicate larger values for PT. Points are shown for each spectral mode, with size weighted by a heat flux $Q_\alpha^{g}$ in PT, such that larger points are linked to larger heat fluxes. Here the label $g$ denotes the type of heat flux where the corresponding quantity enters in the equation, i.e. convective, conductive, or total heat flux. NT and PT were interpolated to common $k_y \rho_s$ values for comparison due to the different local temperatures. The orange line represents a weighted average of modes in the gray area, using the same weights as the point size. The gray shaded area is the same as in  figure~\ref{fig:spectral_analysis}.
    }
    \label{fig:spectral_hf}
\end{figure}

The comparison between NT and PT in the turbulence characterization of the \texttt{GENE-X} simulation (figure \ref{fig:spectral_analysis}) reveals that PT exhibits significantly larger heat fluxes within the range $0.33\lesssim k_y \rho_s \lesssim 1$. Outside this region, the spectra appear to be fairly similar. However, some differences are observed, particularly in the turbulent fluctuations. To better quantify the different contributions, we have extracted all relevant spectral factors of eqs. (\ref{eq:spec_conv}-\ref{eq:spec_cond}) in figure \ref{fig:spectral_hf}. The most pronounced differences are found in fluctuation amplitudes, with NT consistently showing around 10-50\% less fluctuations over the range where the heat fluxes differ. Outside this spectral region, certain modes drive more heat flux in NT, although these modes have a negligible contribution to the total transport. Notably, temperature fluctuations are proportionally stronger suppressed in NT compared to density or electrostatic potential fluctuations. Interestingly, there is no distinct channel where the differences are primarily larger. Temperature fluctuations of ions and electrons, parallel and perpendicular, are similarly suppressed in NT. The strongest effect is observed in the perpendicular electron temperature fluctuations, as the perpendicular conductive electron heat flux is dominant in these cases. No significant differences in phase shifts are observed.

To assess the importance of global effects, local and global simulations must be compared against each other. However, comparing the global simulations in X-point geometry with \texttt{GENE-X} to the local fluxtube simulations with \texttt{GENE} is not a straightforward task. As discussed in Section \ref{sec:gene_model}, several key differences between the models, assumptions, geometry, and numerics make a comparison challenging. In \ref{sec:appendix_gene_nl}, we study the results from linear and non-linear fluxtube simulations with \texttt{GENE} and examine the differences to results from \texttt{GENE-X}.

\subsection{Divertor heat fluxes}

The parallel heat flux on the divertor targets is analyzed in this section. In \texttt{GENE-X}, Dirichlet boundary conditions are applied at the domain boundaries, and a buffer region with numerical dissipation is added for numerical stability \cite{MichelsPhd}. To measure the parallel heat flux, a line is traced from the divertor back upstream to the next toroidal plane, as detailed in Refs. \cite{Michels2022, Ulbl2023}. The parallel heat flux is expected to follow an empirical, so-called Eich-profile, \cite{Eich2013},
\begin{align}
    q_{||}^\mathrm{target}(r) = \frac{q_0}{2} \exp\left(\left(\frac{S}{2\lambda_q}\right)^2-\frac{r}{\lambda_q}\right) \mathrm{erfc}\left(\frac{S}{2\lambda_q}-\frac{r}{S}\right) + q_\mathrm{BG},\label{eq:eich}
\end{align}
where the parameters are background heat flux $q_\mathrm{BG}$, peak heat flux $q_0$, SOL width (or fall-off length) $\lambda_q$ and spreading factor $S$. The argument $r=R-R_\mathrm{sep}$ represents the mapped distance at the OMP to the separatrix. To construct the mapping from the target to the OMP, we use the total poloidal flux $\psi_\mathrm{pol} = \psi_\mathrm{pol}^\mathrm{equi} + 2\pi R \left<A_{1,||}\right>_{t,\varphi}$, which takes into account the mean contributions from the perturbed parallel electromagnetic vector potential.

\begin{figure}
    \centering
    \includegraphics[width=0.9\textwidth]{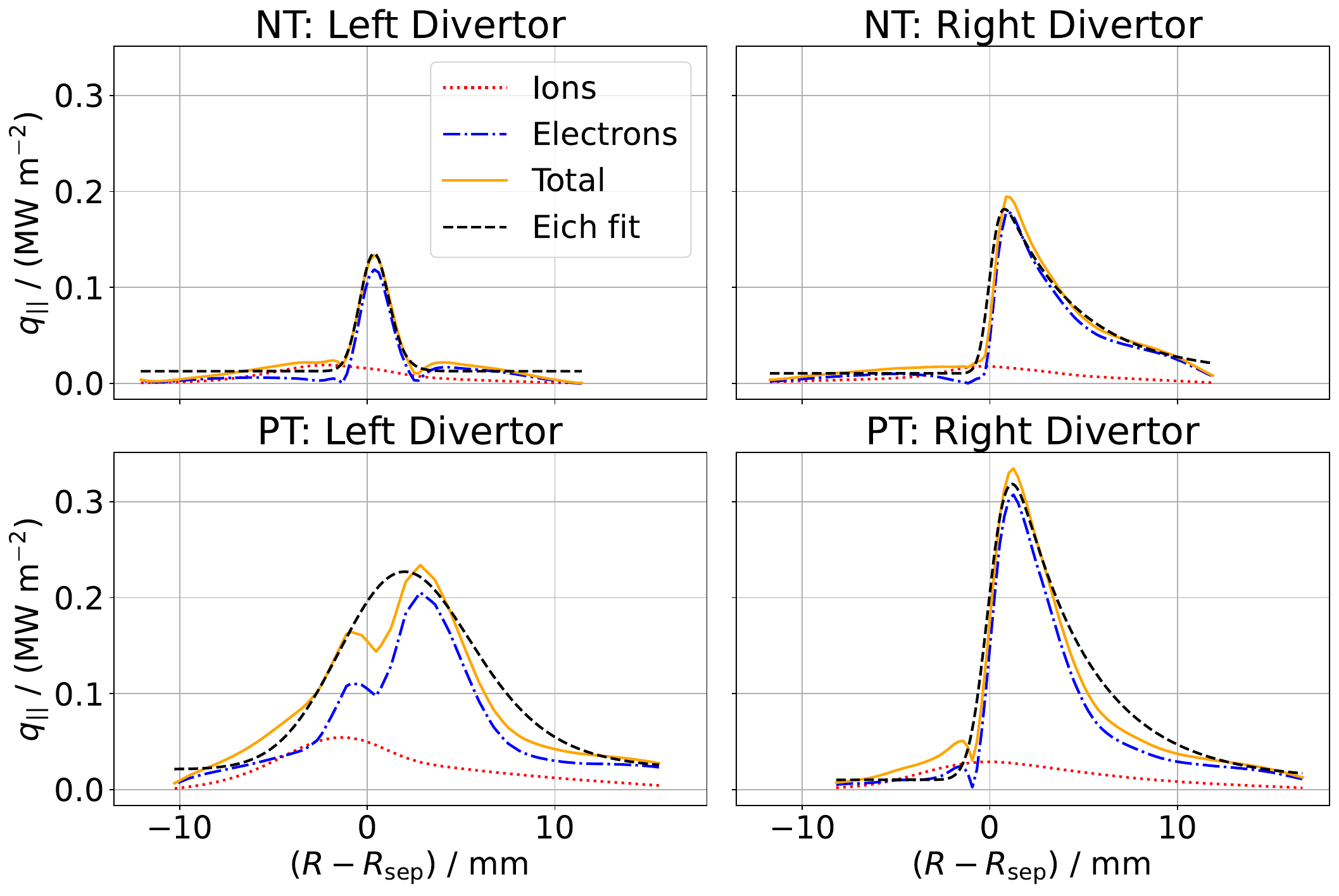}
    \caption{Comparison of parallel heat fluxes on the divertor target plates in the NT and PT \texttt{GENE-X} simulations. The parallel heat flux $q_{||}$ is projected perpendicular to the target by accounting for the incidence angle $\alpha_B$ via multiplication with $\sin(\alpha_B)$. In these cases, the incidence angles are $\alpha_B^\mathrm{NT, left}\approx 1.9^{\circ}-2.7^{\circ}$, $\alpha_B^\mathrm{NT, right}\approx 3.1^{\circ}-3.8^{\circ}$, $\alpha_B^\mathrm{PT, left}\approx 4.7^{\circ}-4.9^{\circ}$, $\alpha_B^\mathrm{PT, right}\approx 2.3^{\circ}-2.7^{\circ}$. The colored lines represent the ion, electron, and total heat flux components, while the black line shows the Eich fit of the total heat flux. The x-axis represents the distance to the strike line on the target plate, mapped back to the outboard midplane (OMP).
    }
    \label{fig:div_hf}
\end{figure}

Measured parallel heat flux profiles in the \texttt{GENE-X} simulations are shown in figure \ref{fig:div_hf}, for both left (inner) and right (outer) divertor targets. Additionally, a fit to the Eich function~(\ref{eq:eich}), as explained in detail in Ref. \cite{TCV-X21}, is shown. We observe that the parallel heat fluxes can be well represented by the Eich function (\ref{eq:eich}). The peak heat flux is mostly carried by the electrons, while the ions provide a broader background and strong transport to the private flux region \cite{Brida2025}. An exception to this observation is the left divertor in the PT simulation, which deviates significantly from the Eich function (\ref{eq:eich}) and a dominant spreading $S$ is observed in that case. The fit parameters for $\lambda_q$ and $S$ are given in the first two rows of table \ref{tab:lambdaq}.

From the analysis of the Eich fit results, we find similar $\lambda_q$ values for NT and PT in the right/outer target, while at the left/inner target, $\lambda_q$ is significantly smaller in NT. Previous modeling results in TCV using fluid models \cite{Muscente2023, Lim2023} show a trend of increasing $\lambda_q$with increasing triangularity for the right/outer target, also observed in experimental configurations \cite{Faitsch2018}. Our simulation results do not directly contradict these findings, particularly for the small triangularities in the cases analyzed here, where similar $\lambda_q$ values at the outer target are realistic. More recent experimental studies have also found similar $\lambda_q$ values in NT and PT in certain cases, considering Langmuir probe data \cite{Fevrier2024}. At the inner target, our simulations follow the experimentally observed trend of smaller SOL width for longer connection lengths of the target to the OMP \cite{Faitsch2018}. Calculating the ratios $\lambda_q^\mathrm{in} / \lambda_q^\mathrm{out}$, we obtain approximately 0.11 for NT and approximately 0.7 for PT. Compared to the \enquote{Goldston} factor $G_s = (1-\delta)/(1+\delta)$ derived via the heuristic drift-based model \cite{Goldston2012}, we find a close match only in PT, $G_s^\mathrm{PT}\approx0.62$, while in NT no match is found, $G_s^\mathrm{NT}\approx1.64$. This trend has also been found in experimental studies \cite{Faitsch2018}. We note that using the top, bottom, or average triangularity does not alter this result in a significant way.

Due to the significant spreading and different geometry of the SOL and divertor, we also analyze the integrated SOL widths. These are given by $\lambda_\mathrm{int} = \lambda_q + 1.64 S$ \cite{Makowski2012} and the width on the target $\lambda_\mathrm{int}^\mathrm{target} = f_x \lambda_\mathrm{int}$ \cite{Faitsch2018}, where $f_x = B_\theta^\mathrm{OMP} B_\varphi^\mathrm{target} / (B_\theta^\mathrm{target} B_\varphi^\mathrm{OMP})$ is the poloidal flux expansion \cite{Theiler2017}. The width $\lambda_\mathrm{int}^\mathrm{target}$ is the quantity directly affecting the wetted area in the divertor \cite{Faitsch2018}. These quantities are given in the last three rows of table \ref{tab:lambdaq}. Firstly, the integrated width takes into account the parallel heat flux spreading, which is significantly larger in PT than in NT. Interestingly, this result has also been found in recent experimental studies over multiple different configurations of NT/PT geometries \cite{Fevrier2024}. In contrast, previous studies using fluid models have not reported a difference in the spreading factor between NT and PT \cite{Muscente2023, Lim2023}. Secondly, due to stronger flux expansion in the LFS-SOL in PT, the integrated width on the right target is around 50\% larger in PT. Conversely, the flux expansion is larger in the LFS-SOL in NT, but the integrated width on the left target is around 3 times smaller in NT.

\begin{table}[]
\centering
    \begin{tabular}{|l||r|r||r|r|}
    \hline
      & \textbf{NT Left} & \textbf{PT Left} & \textbf{NT Right} & \textbf{PT Right} \\\hline
      $\lambda_q$ / mm & 0.46 & 2.76  & 3.90 & 3.93 \\\hline
      $S$ / mm & 0.96 & 4.26 & 0.64 & 1.16 \\\hline
      $\lambda_\mathrm{int}$ / mm & 2.0 & 9.8 & 5.0 & 5.8 \\\hline
      $f_x$ & 2.44 & 1.25 & 1.73 & 2.30 \\\hline
      $\lambda_\mathrm{int}^\mathrm{target}$ / mm & 5.0 & 12.2 & 8.6 & 13.4 \\\hline
    \end{tabular}
    \caption{Parallel heat flux characteristics in the NT and PT \texttt{GENE-X} simulations. A comparison of measured parameters, including SOL width $\lambda_q$ and spreading $S$ from the Eich fit, integrated width $\lambda_\mathrm{int}$, flux expansion $f_x$ and integrated width at the target plates $\lambda^\mathrm{target}_\mathrm{int}$.
    }
    \label{tab:lambdaq}
\end{table}

\section{Summary and discussion}

For the first time, we conducted an extensive multi-fidelity study of fluid and gyrokinetic turbulence in diverted negative (NT) and positive triangularity (PT) geometries. First-principles simulations using the \texttt{GENE-X} code show that in comparable NT and PT geometries, similar profiles are achieved in the saturated turbulent state. However, the turbulent ExB heat flux is substantially reduced by more than 50\% in NT, resulting in an improvement in energy confinement of approximately 67\%. In contrast, simulations with the drift-reduced fluid turbulence code \texttt{GRILLIX}, including a neutrals model and sheath boundary conditions, were unable to capture the improved confinement observed in NT, suggesting that the enhanced confinement is due to the impact of NT on trapped electron modes (TEMs). Local gyrokinetic simulations with \texttt{GENE} confirm that TEMs are the dominant micro-instability present in the system. The heat flux reduction in NT is attributed to smaller fluctuation levels due to the reduced TEM drive. Analysis of the parallel heat flux on the divertor targets reveals a distinct effect on heat flux spreading. In NT, the spreading $S$ is consistently smaller at both targets, while the SOL widths $\lambda_q$ are either smaller or similar compared to PT.

We linked the improvement of confinement to the presence of TEMs in the system. Linear fluxtube simulations with the \texttt{GENE} code revealed that TEMs are linearly unstable. This result is consistent with an in-depth turbulence characterization of the saturated non-linear turbulent state resulting from \texttt{GENE-X}. The electrostatic ExB heat flux is primarily driven by electrons, with perpendicular temperature fluctuations playing a significant role. The density and ion temperature phase shifts exhibit drift-wave characteristics, while the electron temperature phase shifts exhibit interchange-like behavior. Furthermore, broadband turbulence propagating in the electron diamagnetic direction was observed, with a frequency close to the linear TEM frequency. In contrast, \texttt{GRILLIX} shows drift wave turbulence based on these criteria, while linear fluxtube simulations from \texttt{GENE} suggest TEM turbulence based on \texttt{GRILLIX} profiles. The differences between NT and PT in all spectral heat flux contributions were studied, revealing a correlation between the reduction in turbulent heat flux in NT and decreased turbulent fluctuations. Overall, we conclude that the observed increase in confinement time in NT is attributed to a reduction of the TEM-driven turbulence, an effect that is only present in the gyrokinetic models.

We analyzed the scrape-off layer (SOL) width $\lambda_q$ with \texttt{GENE-X} in NT and PT. A key finding of this study is that the heat flux spreading $S$ and $\lambda_q$ are differently affected in NT and PT. We reproduced experimental observations that the spreading is smaller in NT \cite{Fevrier2024}.
We found similar values of $\lambda_q$ at the right/outer target, while at the left/inner target NT shows a much narrower SOL width. The outer target result is similar to experimental findings in Ref. \cite{Fevrier2024}, while the inner target result is consistent with the increase of the parallel connection length between target and OMP \cite{Faitsch2018}. The ratio between inner and outer target SOL widths matches predictions by the Goldston model \cite{Goldston2012} only in PT. The integrated width $\lambda_\mathrm{int}$ is wider for both targets in PT. Taking into account flux expansion, the width is around 50\% larger on the outer target and 3 times larger on the inner target in PT. It is worth noting that the flux expansion can be varied significantly without changing, in leading order, the parallel heat flux onto the target \cite{Theiler2017}. A future gyrokinetic study of varying triangularity for the same divertor geometry is required in this context. A fair comparison of the SOL width should consider more competitive scenarios for a reactor, such as NT L-mode vs. PT H-mode.

Previous studies have found that the beneficial effect of NT on TEM turbulence is due to the stabilization of the toroidal precession drift resonance \cite{Marinoni2009}. This is supported by analytical calculations, as shown in Ref. \cite{Garbet2024}. However, the question remains as to whether the turbulence stabilization observed in the global non-linear \texttt{GENE-X} simulations in this work can be directly attributed to this effect. Linear \texttt{GENE} simulations indicate that the beneficial effect comes from the edge, similar to findings in Ref. \cite{Mariani2024b}. Nevertheless, it is unclear whether this is due to different boundary conditions at the separatrix or modified physics in the SOL. Global effects are anticipated to be important due to the steep profile gradients and small gradient length scales at the edge.

Future work should include additional code validation in NT geometry, such as simulations on AUG, DIII-D, and JET, as well as the exploration of different turbulent regimes, including ion temperature gradient (ITG) turbulence \cite{Merlo2023a, Merlo2023b, Balestri2024}. Additionally, comparing codes like \texttt{GENE} global \cite{Goerler2011} and \texttt{GENE-X} would provide valuable insights into the underlying physics. However, this is a complex task due to the different approaches in physics models, numerics and geometry. Along these lines, further model extensions of the \texttt{GENE-X} code, such as higher-order FLR effects, need to be implemented and their effect on turbulence studied. Reactor-relevant scenarios, in particular NT L-mode vs. PT H-mode, are an interesting next application case. In this context, the influence of NT on electromagnetic instabilities, such as kinetic ballooning modes (KBM) \cite{Parisi2024}, needs to be investigated in global simulations.

\section*{Acknowledgements}

The authors would like to thank S. Coda, A. Balestri and the TCV team for providing the magnetic equilibria. Further, the authors thank W. Zholobenko, B. Frei, D. Brida, M. Faitsch and the EUROfusion TSVV task 2 and task 4 teams for useful discussions. The simulations presented herein were carried out on the Marconi supercomputer at CINECA and the Cobra and Raven supercomputers at MPCDF.

This work has been carried out within the framework of the EUROfusion Consortium, funded by the European Union via the Euratom Research and Training Programme (Grant Agreement No 101052200 — EUROfusion). Views and opinions expressed are however those of the author(s) only and do not necessarily reflect those of the European Union or the European Commission. Neither the European Union nor the European Commission can be held responsible for them.

\appendix
\section{Comparison of local \texttt{GENE} against global \texttt{GENE-X} simulations}\label{sec:appendix_gene_nl}
\setcounter{section}{1}

\subsection{Nonlinear simulation setup}

For the nonlinear simulations, we use \texttt{GENE} with $(256\times32\times32\times32\times18)$ points in $(k_{x}\times k_{y}\times z\times v_{\|}\times \mu)$. For the outermost radial position in positive triangularity, the strong magnetic shear combined with increased parallel variation of the magnetic geometry suggested increasing the radial ($x$) and parallel ($z$) resolutions to 512 and 64 points respectively. The perpendicular box size $(x\times y)$ is around $(120\rho_\mathrm{s}\times120\rho_\mathrm{s})$. The code slightly adjusts these to match machine size and radial box size quantization requirements. We remark that these numbers are \enquote{nominal} in that they are modified by flux expansion and compression. As an example, this corresponds to a box width of about 6 cm at the OMP at $\rho_{\mathrm{tor}}=0.95$
in positive triangularity, compared to a minor radius of about 25 cm. All other settings are as described in section \ref{sec:gene_sim}. The non-linear simulations were carried out on the Raven supercomputer at MPCDF. The computational cost is around 300 kCPUh in total for all cases.

\subsection{Linear simulation results}

Comparing the results of the linear fluxtube \texttt{GENE} simulations (figure \ref{fig:gene}) against \texttt{GENE-X} reveals a somewhat discrepant picture. The latter shows differences in NT and PT that depend strongly on the radial location that is analyzed, whereas the former shows overall smaller heat fluxes in NT over the entire radial domain. For example, at $\rho_\mathrm{pol}=0.95$ ($\rho_\mathrm{tor}\approx0.88$), \texttt{GENE} predicts similar growth rates for very different gradient lengths, while \texttt{GENE-X} shows a 50\% decrease in the spectral heat fluxes in NT locally (figure \ref{fig:spectral_hf}).

This discrepancy can potentially be explained by multiple effects. Radially outwards $\rho_\mathrm{pol}=0.98$ ($\rho_\mathrm{tor}\approx0.95$) triangularity increases, correlating to a strong suppression of linear growth rates in NT (figure \ref{fig:gene}). Additionally, magnetic shear increases significantly $\hat{s}\gtrsim 5$ approaching the separatrix. Both parameters have been found to affect linear growth rates in TEM-dominated cases \cite{Merlo2023a}. Global effects can contribute to the observed qualitative difference between the global \texttt{GENE-X} and fluxtube \texttt{GENE} results. Previous studies have found that non-local effects reduce heat fluxes more strongly in NT compared to local simulations \cite{Merlo2021}. The plasma profiles are different which might overshadow or enhance some of these effects. The next section discusses the influence of plasma profiles and global effects.


\subsection{Nonlinear simulation results}

\begin{figure}
    \centering
    \includegraphics[width=0.7\textwidth]{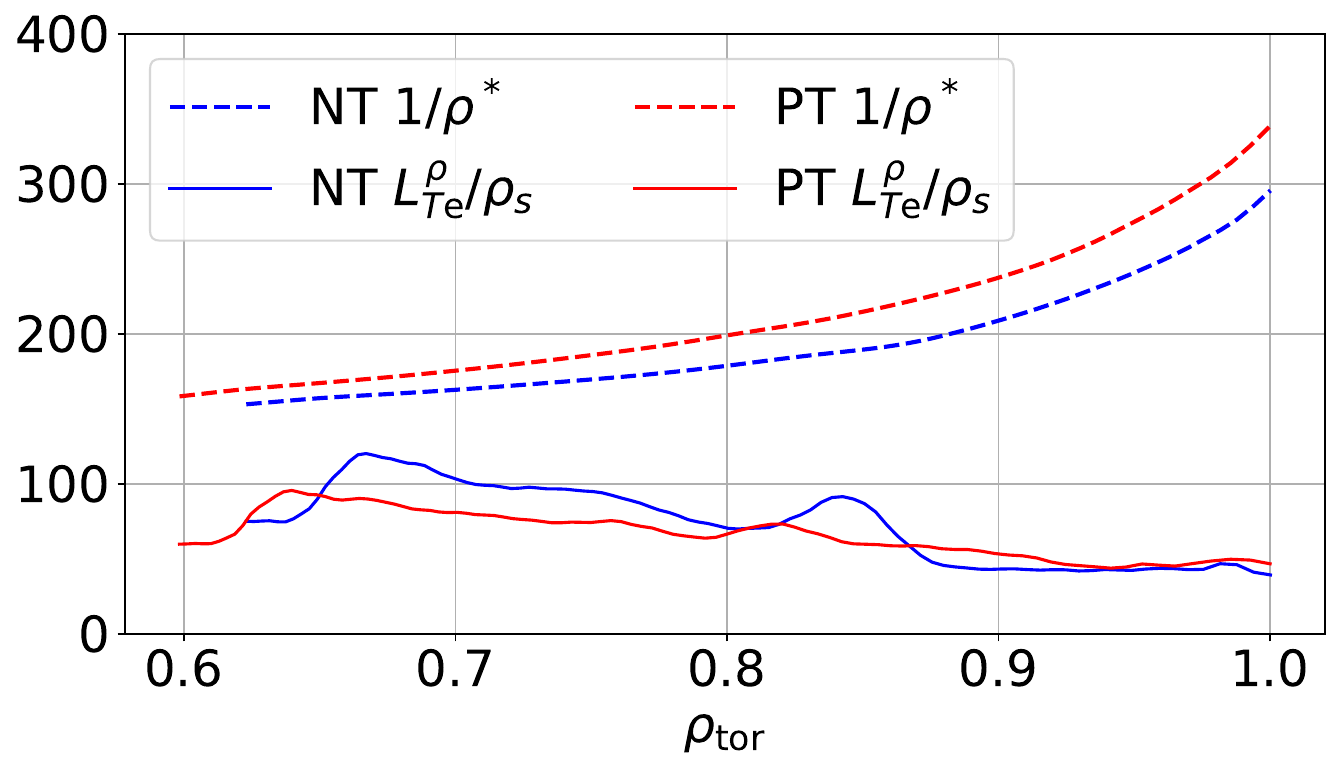}
    \caption{Profiles of inverse normalized scale length $1/\rho^*$ and inverse normalized gradient length $L_{T\mathrm{e}}^\rho/\rho_s$ for NT and PT (see figure \ref{sec:gene_sim} for definition).
    }
    \label{fig:gene_grad}
\end{figure}

Gyrokinetic fluxtube simulations with the \texttt{GENE} code were conducted for radial locations $\rho_\mathrm{tor}\in[0.8, 0.88, 0.95]$. The results for integral heat fluxes and diffusivities are presented in table \ref{tab:gene_nl}. We observe that the integral heat fluxes are comparable between NT and PT at the inner locations $\rho_\mathrm{tor}\in[0.8, 0.88]$, while at the outermost flux surface considered ($\rho_\mathrm{tor}=0.95$), NT has around 43\% less heat flux. This trend is consistent for both ions and electrons, as well as for the particle flux (not shown). The disparate behavior between NT and PT can be attributed to the different profiles and gradients in both cases. The heat flux scales with $Q\sim T^{5/2}$, resulting in more heat flux being driven due to the larger temperatures in NT. This profile effect diminishes towards the edge, as the profiles approach each other in NT and PT as $\rho_\mathrm{tor}\to1$. Additionally, the turbulent diffusivities are consistently smaller in NT, particularly at the outside radial locations $\rho_\mathrm{tor}\in[0.88, 0.95]$. This is attributed to the overall steeper profiles in NT. The main discrepancy in the non-linear fluxtube results concerns the different integral heat fluxes at different radial locations, which appear to differ from the global \texttt{GENE-X} simulations. It is likely that global effects play a significant role, despite approaching the edge of the confined region. This is evident in figure \ref{fig:gene_grad}, where the standard measure for global effects $1/\rho^*$ is compared to the normalized gradient lengths $L_{T\mathrm{e}}/\rho_s$. Due to the steep gradients present in the profiles, global effects are expected to become increasingly important as $\rho_\mathrm{tor}\to1$ in this case.

Extracting the precise stabilization mechanism observed in the non-linear fluxtube simulations in NT is challenging due to the complex interplay of multiple factors. A multi-parametric dependency on profiles, gradients, safety factor, shear, and triangularity is anticipated to contribute to the observed behavior. To fully understand the potential influence of these factors, a comprehensive study of all possible combinations of these parameters, including both NT and PT equilibria, is required. This investigation is left for future work.

\begin{table}[]
\centering
    \begin{tabular}{|l||r|r||r|r||r|r|}
    \hline
      & \textbf{NT 0.8} & \textbf{PT 0.8} & \textbf{NT 0.88} & \textbf{PT 0.88}  & \textbf{NT 0.95} & \textbf{PT 0.95}\\\hline\hline
      $Q_\mathrm{i}^\mathrm{es}$ / kW & 140 & 130 & 420 & 470 & 190 & 300\\\hline
      $Q_\mathrm{e}^\mathrm{es}$ / kW & 230 & 190 & 600 & 510 & 250 & 470\\\hline\hline
      $\chi_\mathrm{i}^\mathrm{es}$ / (m$^2$ s$^{-1}$) & 4.19 & 4.28 & 9.93 & 15.15 & 3.96 & 10.43\\\hline
      $\chi_\mathrm{e}^\mathrm{es}$ / (m$^2$ s$^{-1}$) & 4.84 & 5.04 & 9.99 & 13.44 & 4.95 & 10.85\\\hline
    \end{tabular}
    \caption{Results of non-linear fluxtube simulations with the \texttt{GENE} code at radial locations $\rho_\mathrm{tor}\in[0.8, 0.88, 0.95]$ for PT and NT. Shown are ion and electron electrostatic integral heat fluxes $Q^\mathrm{es}$ and diffusivities $\chi^\mathrm{es}$. Electromagnetic contributions are negligible.}
    \label{tab:gene_nl}
\end{table}

\printbibliography

\end{document}